\documentclass[12pt]{article}
\usepackage{float}\restylefloat{figure}
\usepackage{graphicx}
\usepackage{epsfig}
\usepackage{subfigure}
\usepackage{rotating}
\usepackage{caption}
\usepackage{epstopdf}
\usepackage{amssymb}

\def\bm#1{\mbox{\boldmath $#1$}}
\def\beq{\begin{equation}}
\def\eeq{\end{equation}}
\def\t2{\mbox{  }}
\def\rst1{\mbox{ }}

\begin{document} 
\title{\normalsize {\bf Structure and thermodynamics of a ferrofluid bilayer.}}
\author{\normalsize  Carlos Alvarez$^{1,2,3}$, Martial Mazars$^{1}$\footnote{author for correspondance : martial.mazars@th.u-psud.fr}\ \  and Jean-Jacques Weis$^{1}$ \\
\small  $^{1}$ \it Laboratoire de Physique Th\'eorique (UMR 8627),\\
\small \it Universit\'e de Paris Sud XI, B\^atiment 210, 91405 Orsay Cedex,
FRANCE\\[0.1in]
\small $^{2}$ \it Departamento de Fisica, \\
\small \it Universidad de Los Andes, Carrera 1E\# 18-10, Bogot\'a, COLOMBIA\\[0.1in]
\small  $^{3}$ \it Laboratoire de Physique Th\'eorique et Mod\`eles Statistiques (UMR 8626),\\
\small \it Universit\'e de Paris Sud XI, B\^atiment 100, 91405 Orsay Cedex,
FRANCE}
\maketitle
\hfill\small\hspace{1.0in} Preprint  L.P.T.-Orsay : 08-21\\ 
\begin{center}{\bf Abstract}\end{center}
We present extensive Monte Carlo simulations for the thermodynamic and structural properties of a planar bilayer of dipolar hard spheres for a wide range of densities, dipole moments and layer separations. Expressions for the stress and pressure tensors of the bilayer system are derived. For all thermodynamic states considered the interlayer energy is shown to be attractive and much smaller than the intralayer contribution to the energy. It vanishes at layer separations of the order of two hard sphere diameters. The normal pressure is negative and decays as a function of layer separation  $h$ as $-1/h^5$.  Intralayer and interlayer pair distribution functions and angular correlation functions are presented. Despite the weak interlayer energy strong positional and orientational correlations exist between particles in the two layers.
\newpage
{\centering
\section{INTRODUCTION}\label{Intro}
}
Dipolar interactions play a significant role in determining the structural, magnetic or rheological properties of a variety of quasi two-dimensional (2D) systems (monolayers, multilayers, thin films) including suspensions of colloidal particles at an air-water interface, adsorbed amphiphilic molecules, lipid bilayers, ultrathin magnetic films etc.. (see e.\ g.\ ref. \cite{Weis:03} and references therein). In most of these systems the properties and phase behavior result, though, from an interplay of the dipolar interaction with competing interactions, as for instance, the hydrocarbon chain tails or water mediated interactions in lipid bilayers \cite{Saiz:02,Scott:02}, or exchange interaction and magneto crystalline anisotropy in thin magnetic films \cite{DeBell:00}. Although simulations taking into account full atomic details have been performed in the past (generally computationally costly) for these kinds of systems (see e.g. ref.\cite{Pertsi:07} and references therein)  we believe that a study of a purely dipolar bilayer system is of interest in its own right providing unbiased insight into the role of the dipolar interaction. The experimental system which perhaps comes closest to the pure dipolar system is the ferrofluid system. In effect, association into chains, rings, branched structures or stripes has been demonstrated in recent
experiments on strongly interacting ($\mbox{Fe}_3\mbox{O}_4$) ferrofluids \cite{Klokke:04,Klokke:06a,Klokke:06b} and comparison with simulation results presenting similar structures is more than suggestive that the dipolar hard sphere (DHS) system is a fair representation of these types of ferrofluid.\\
Extensive Monte Carlo (MC) simulation  and theoretical results for the self organization of quasi 2D DHS are already available for the monolayer system both with and without an external field \cite{Weis:02,Tavare:02,Tavare:06,Weis:02a,Satoh:96,Lomba:00,Weis:05,Duncan:04}.\\
The purpose of the present paper is to extent these results to a symmetric planar bilayer the main interest, evidently, being to probe the effect of the interlayer interaction on particle organization.\\ 
In Sect.\ \ref{Model} we define the bilayer model and give details of the numerical simulation methods we use. The next section gives expressions for the energy, stress tensor and correlation functions of the bilayer system. Sect.\ \ref{Results} contains the simulation results for the thermodynamic and structural properties. A summary is given in the last section. The three appendices A-C provide expressions for the Ewald sums of energy (A), pressure and forces (C) and a derivation of the microscopic stress tensor of the bilayer (B).

{\centering
\section{MODEL AND NUMERICAL METHODS}\label{Model}
}
The systems consist of $N=2N_0$ particles with permanent point dipole moment $\mu$ interacting via hard sphere and dipolar potentials. Particles are evenly distributed among two layers $L_1$ and $L_2$ separated by a distance $h$, each layer being rectangular with sides $L_x$ and $L_y$ ; $A=L_xL_y$ is the surface area of the layers. Periodic boundary conditions (p.b.c.), with spatial periodicities $L_x$ and $L_y$, are applied in the directions $x$ and $y$ parallel to the layers, but no p.b.c.  are taken in the third direction $z$. Particle positions are constrained to lie in the layers but dipole moments can orient in full 3D space. The interaction potential between the particles is pairwise additive and is  represented as
\begin{equation}
\label{pot1}
\Phi(\bm{r}_{ij},\bm{\mu}_i,\bm{\mu}_j)=\left\{\begin{array}{ll} 
\displaystyle \infty &\displaystyle \mbox{for   } r_{ij} <\sigma\\
&\\
\displaystyle \frac{1}{r_{ij}^3}\mbox{\Large{[}}\bm{\mu}_i\cdot\bm{\mu}_j - 3(\bm{\mu}_i\cdot\hat{\bm{r}}_{ij})(\bm{\mu}_j\cdot\hat{\bm{r}}_{ij})\mbox{\Large{]}} &\displaystyle \mbox{for    } r_{ij} >\sigma
\end{array}\right.
\end{equation}
where $\sigma=1$ is the hard sphere diameter taken as unit length, $\bm{\mu}_i$ the dipole moment of particle $i$ and $\hat{\bm{r}}_{ij}=\bm{r}_{ij}/r_{ij}$ the unit bond vector between particles $i$ and $j$. In the following, we will use the notations
\begin{equation}
\label{pot2}
\bm{r}_{ij}=\bm{s}_{ij}+z_{ij}\rst1\hat{\bm{e}}_z \mbox{     and     } \bm{\mu}_i=\mu \hat{\bm{\mu}}_i
\end{equation}
where $\hat{\bm{e}}_z$ is the unit vector perpendicular to the layers and $\hat{\bm{\mu}}_i$ a unit vector in the direction of dipole moment $i$.\\ 
Only  surface separation $h > 1$ which avoid hard core interactions between the layers have been considered. A few simulation results for  $h > 1$ have been presented previously by one of us \cite{Weis:03}.\\ 
Monte Carlo (MC) simulations have been performed in the canonical (NVT) ensemble with system sizes comprising $N=1024-3200$ particles. The total number of MC cycles varied from $0.2 \times 10^6$ to $2 \times 10^6$, depending on density and dipole moment, each cycle consisting of displacement and rotation of the $N$ particles. The amplitude of the trial moves was chosen to obtain  acceptance ratios between 30 and 50\% for each thermodynamic state. No exchange of particles between layers $L_1$ and $L_2$ is allowed.\\ 
Reduced quantities for surface area, $A^* = A/ \sigma^2$, surface density $\rho^*=\rho \sigma^2= N_0 /A \sigma^2$, and dipole moment $\mu^* = ( \mu^2/kT \sigma^3)^{1/2}$ will be used throughout the paper. For notational convenience the stars will be dropped.

{\centering \section{THERMODYNAMICAL AND STRUCTURAL QUANTITIES}\label{Thermo} }
{\centering \subsection{Energy} }
In our model the energy of the bilayer is entirely given by the dipolar contribution which we split into an intralayer contribution, $U^{intra}$, and an interlayer contribution, $U^{inter}$, as
\begin{equation}
\displaystyle U_{dd} = U^{intra}+U^{inter}.
\label{ene1}
\end{equation}
These are computed using the Ewald method \cite{Heyes:94,Grzybo:00,Weis:03,Weis:05a} ; the relevant expressions for $U^{intra}$ and  $U^{inter}$ are given in Appendix A.\\  
For bulk systems with slab geometry where periodicity applies only in two spatial directions, say $L_x$ and $L_y$, the Ewald sums are computationally costly due to the appearance in the reciprocal space term of a double sum over the distance $z_{ij}$ in the bounded direction of particles $i$ and $j$ \cite{Heyes:94, Grzybo:00}. As in the present case the distance $z_{ij}$ between two particles will be constant, the corresponding sums can be reduced to order $N$ \cite{Weis:03} similar to the cases of  Coulomb \cite{Weis:01,Mazars:05} or Yukawa \cite{Mazars:07} potentials.\\
One can note  that the 3D bilayer system can be mapped onto a two-component monolayer system by considering the particles in the two layers as distinct species\cite{Valtchinov:97}. For most of the thermodynamical and structural quantities, both approaches are equivalent ; for instance, in the two-component monolayer, $U^{inter}$ is the total interaction between particles belonging to different species (different layers). As outlined in the next subsection and in Appendix B, for pressures and stresses such a mapping is slightly less straightforward.

{\centering
\subsection{Surface stress tensor and normal pressure}
}
Characterizing the pressure in the bilayer system needs some care. In particular, since the particles are constrained to belong to layers $L_1$ and $L_2$, some degrees of freedom of the particles are frozen by the geometrical features of the system. These constraints have obviously an influence on the flux of momentum per unit area in the system and therefore affect the stress tensor. For the sake of definitness a full derivation of the stress tensor from the lagrangian function of the bilayer system is given in Appendix B.\\
As for systems with slab geometry or interfaces \cite{Widom:82}, the stress tensor is decomposed into lateral and normal components. According to Eq.(\ref{BB13}-\ref{BB15}), the lateral component to the pressure tensor is given by
\begin{equation}
\label{press1}
\begin{array}{ll} 
\displaystyle \Pi_{T}= 2\rho kT &\displaystyle- \frac{1}{4A}\Big<\sum_{i\in L_1}\sum_{j\in L_1, j\neq i}\bm{s}_{ij}.\bm{\nabla}_{i}\Phi(\bm{s}_{ij},0)\Big>\\
&\\
&\displaystyle- \frac{1}{4A}\Big<\sum_{i\in L_2}\sum_{j\in L_2, j\neq i}\bm{s}_{ij}.\bm{\nabla}_{i}\Phi(\bm{s}_{ij},0)\Big>\\
&\\
&\displaystyle -\frac{1}{2A}\Big<\sum_{i\in L_1}\sum_{j\in L_2} \bm{s}_{ij}.\bm{\nabla}_{i}\Phi(\bm{s}_{ij},h)\Big>
\end{array}
\end{equation}
where $\Phi(\bm{s}_{ij},h)$ is the pair potential.\\
From the point of view of mapping the bilayer system onto a two-component monolayer system, the lateral pressure $\Pi_{T}$ in the bilayer, defined in Eq.(\ref{press1}) through Eqs.(\ref{BB12}-\ref{BB15}), corresponds to the pressure of the 2D, two-component monolayer system. In solid surface physics, $\Pi_T$ is related to the surface stress $\tilde{\eta}$ by $\Pi_T=-\tilde{\eta}$ (cf. Eq.(\ref{BB15})), and for fluids confined in slab geometry $\Pi_T$ is related to the lateral pressure $P_T(z)$ by
\begin{center}
$\displaystyle \Pi_T = \int \mbox{ } dz\mbox{ }P_T(z).$
\end{center}
$\Pi_T$ can be composed into ideal, hard sphere (HS), and dipolar contributions
\begin{equation}
\label{press2}
\displaystyle \Pi_{T}= 2\rho kT+2\Pi^{(HS)}_{T}+\Pi^{(HS)}_{T,inter}+\Pi^{(dd)}_{T}
\end{equation}
where the dipolar part $\Pi^{(dd)}_{T}$ is obtained from Eq.(\ref{pot1}) and the relation 
\begin{equation}
\label{press3}
\begin{array}{ll} 
\displaystyle s_{ij}^{\beta}\nabla_{i}^{\alpha}\Phi^{(dd)}(\bm{s}_{ij},h)&\displaystyle = 3\frac{s_{ij}^{\beta} s_{ij}^{\alpha}}{ (s_{ij}^{2}+h^2)^{5/2}}\left [ \bm{\mu}_i .\bm{\mu}_j -5\frac{(\bm{\mu}_i.\bm{s}_{ij}+\mu_i^z h) (\bm{\mu}_j.\bm{s}_{ij}+\mu_j^z h)}{s_{ij}^{2}+h^2}\right ] \\
&\\
&\displaystyle-3 \frac{s_{ij}^{\beta}}{ (s_{ij}^{2}+h^2)^{5/2}}\left [ (\bm{\mu}_i.\bm{s}_{ij}+\mu_i^z h)\mu_j^{\alpha}+(\bm{\mu}_j.\bm{s}_{ij}+\mu_j^z h)\mu_i^{\alpha}\right ]
\end{array}
\end{equation}
(see  Eq.\ (\ref{BB14}) of Appendix B). $\Pi^{(dd)}_{T}$ contains both intralayer contributions of layers $L_1$ and $L_2$ and the interlayer contribution; thus, for $h\rightarrow \infty$, $\Pi^{(dd)}_{T}$ is twice the dipolar contribution to the 2D pressure of a monolayer. The dipolar interlayer contribution to $\Pi_T$ is given by the last contribution in the right-hand side (r.h.s.) of Eq.(\ref{press1}) ; this contribution becomes very small as soon as $h\gtrsim 2$.\\
The hard sphere contributions $\Pi_{T}^{(HS)}$ and $\Pi^{(HS)}_{T,inter}$, are computed from the contact values of the intralayer, $g_{intra}^{000}(\sigma)$, and interlayer, $g_{inter}^{000}$, pair distribution functions, defined below, as
\begin{equation}
\label{press4}
\left \{ \begin{array}{ll} 
\displaystyle \Pi_{T}^{(HS)} &\displaystyle =\frac{\pi}{2}\rho^2 kT g_{intra}^{000}(\sigma)\\
&\\
\displaystyle \Pi_{T,inter}^{(HS)} &\displaystyle =\frac{\pi}{2}(2\rho)^2 kT g_{inter}^{000}\left(\sigma\sqrt{1-\frac{h^2}{\sigma^2}}\right)\\
\end{array}
\right.
\end{equation}
As in the present work, $h>1$ in all computations, we always have $\Pi_{T,inter}^{(HS)}=0$. In the limit  $h\rightarrow\infty$ and $\mu\rightarrow 0$, $\Pi_{T}^{(HS)}$ equals the excess contribution to the pressure of a monolayer of hard disks with surface density $\rho$. Moreover, for $h\geq 1$ and $\mu=0$, $\Pi_{T}^{(HS)}$ can be approximated quite accurately by available equations of state of hard disks (see e.g. ref.\cite{Santos:95}).\\
The asymptotic behaviour of $\Pi_{T}$ given by Eq.(\ref{press2}) can be understood as follows. In the limit $h\rightarrow \infty$ and $\mu\neq 0$, $\Pi_{T}$, given by Eq.(\ref{press2}), is exactly twice the 2D pressure of a monolayer of DHS with the same $\rho$ and $\mu$. In this limit, if the system is viewed as a two component monolayer system, the two species remain distinct but there will be no interaction between particles belonging to different species. Thus, $\Pi_{T}/2$ is exactly the partial pressure of each component and the bilayer is fully equivalent to a mixture of two kinds of particles confined in a monolayer with HS and dipolar interactions between like particles but no interactions between unlike particles.\\
In the opposite limit $h\rightarrow 0$ and $\mu\neq 0$, the two species become equal and the bilayer system reduces to  a one component monolayer system with a surface density $2\rho$ (provided that $2\rho$ is less than the density at close packing  of hard disks). Obviously, in this limit, the contribution $\Pi_{T,inter}^{(HS)}$ has also to be included in Eq.(\ref{press2}), and $\Pi_{T}$ equals the 2D pressure of a monolayer of dipolar hard disks with a surface density $2\rho$ and same $\mu$. Also, as in this limit particles become indistinguishable,  entropy contributions must be modified accordingly.\\ 
The average normal force by unit area (or normal pressure) is obtained from Eq.(\ref{BB19}) as
\begin{equation}
\label{press5}
\begin{array}{ll} 
\displaystyle P_{zz}&\displaystyle =-\frac{1}{A}\Big <\frac{\partial}{\partial z}\sum_{i\in L_1}\sum_{j\in L_2}\left.\Phi(\bm{s}_{ij},z)\right|_{z=h}\Big > = - \frac{N}{A} \Big < \frac{\partial \beta U^{inter}/N}{\partial h} \Big >\\
&\\
&\displaystyle = P_{zz}^{(dd)}+P_{zz}^{(HS)}
\end{array}
\end{equation}
where $P_{zz}^{(dd)}$ and $P_{zz}^{(HS)}$ denote the contributions from dipolar and HS interactions, respectively. The dipolar parts, $P_{zz}^{(dd)}$ and $\Pi_{T}^{(dd)}$, are computed using Ewald sums, as described in Appendix C. Since in the present work all computations are done with $h>1$ one has always $P_{zz}^{(HS)}=0$. The HS repulsion does, however, contribute to the normal component of the pressure tensor indirectly via the spatial positions of the particles in the  layers. A similar remark applies to the interlayer correlation functions defined below. Eq.(\ref{press5}) agrees with previous derivations for the normal pressure in slab-like geometry \cite{Schofield:82,Walton:85,Klapp:02} or interfaces \cite{Widom:82} . The main difference between Eq.(\ref{press5}) and these  relations is that there is no kinetic (ideal gas) contribution  in Eq.(\ref{press5}), as a consequence of the constraints that apply to the bilayer systems (cf. Eq.(\ref{BB7})). Thus, $P_{zz}$ has to be considered as an average force by unit area normal to the surface rather than a normal pressure.\\
The surface stress tensor is related to the surface free energy par unit area $\gamma$ (or surface tension) by the Shuttleworth equation \cite{Shuttleworth:50}
\begin{equation}
\label{press6}
\displaystyle \eta_{\alpha \beta}=\gamma\delta_{\alpha \beta}+\frac{\partial\gamma}{\partial\epsilon_{\alpha \beta}}
\end{equation}
where $\epsilon_{\alpha \beta}$ is the 2D strain tensor. In fluid
phases, the second contribution in the r.h.s. of Eq.(\ref{press6}) is
null and Eq.(\ref{press6}) reduces to
$\eta_{\alpha\beta}=\gamma\delta_{\alpha \beta}$. This is the case in
most computations done in the present work, except those at high
densities. Since in our computations the surface and the shape of the
layers are kept constant, we do not have access to $\gamma$.

{\centering
\subsection{Correlation functions}
}
The structure of the bilayer system has been characterized, analogously to the  monolayer case \cite{Lomba:00,Weis:02}, by a one particle orientational distribution function of the  dipoles and several pair correlation functions.\\ 
The orientational distribution function $f(\hat{\bm{\mu}})$, measuring the orientation of the  particle dipole moments with respect to the layer normal, is defined from the one-body density as
\begin{equation}
\displaystyle \rho^{(1)}(\bm{r},\hat{\bm{\mu}})=\Big<\sum_{i}\delta(\bm{r}_i-\bm{r})\delta(\hat{\bm{\mu}}_i-\hat{\bm{\mu}})\Big>=\frac{\rho}{4\pi}f(\hat{\bm{\mu}})
\end{equation}
Pair correlation functions are derived from the general definition of the two-body density 
\begin{equation}
\displaystyle \rho^{(2)}(\bm{r},\bm{r}',\hat{\bm{\mu}},\hat{\bm{\mu}}')=\Big<\sum_{i\neq j}^N\delta(\bm{r}_i-\bm{r})\delta(\bm{r}_j-\bm{r}')\delta(\hat{\bm{\mu}}_i-\hat{\bm{\mu}})\delta(\hat{\bm{\mu}}_j-\hat{\bm{\mu}}')\Big>
\end{equation}
where $\hat{\bm{\mu}}$ and $\hat{\bm{\mu}}'$ are unit vectors along the dipole moments. Specifying to intralayer $\rho_{intra}^{(2)}$ and interlayer $\rho_{inter}^{(2)}$ two-body surface densities one has
\begin{equation}
\left\{\begin{array}{ll}
\displaystyle
\rho_{intra}^{(2)}(s,\hat{\bm{\mu}},\hat{\bm{\mu}}')&\displaystyle = \frac{1}{4\pi s}\Big<\sum_{i\in L_{1}}\sum_{j\in L_{1},j\neq i}\delta(s-\mid {\bm s}_{ij}\mid )\delta(\hat{\bm{\mu}}_i-\hat{\bm{\mu}})\delta(\hat{\bm{\mu}}_j-\hat{\bm{\mu}}')\\
&\\
&\displaystyle +\sum_{i\in L_{2}}\sum_{j\in L_{2},j\neq i}\delta(s-\mid {\bm s}_{ij}\mid)\delta(\hat{\bm{\mu}}_i-\hat{\bm{\mu}})\delta(\hat{\bm{\mu}}_j-\hat{\bm{\mu}}')\Big>\\
&\\
\displaystyle \rho_{inter}^{(2)}(s,\hat{\bm{\mu}},\hat{\bm{\mu}}')&\displaystyle =\frac{1}{2\pi s}\Big<\sum_{i\in L_{1}}\sum_{j\in L_{2}}\delta(s-\mid {\bm s}_{ij}\mid)\delta(\hat{\bm{\mu}}_i-\hat{\bm{\mu}})\delta(\hat{\bm{\mu}}_j-\hat{\bm{\mu}}')\Big>
\end{array}
\right. 
\end{equation}
The  intralayer $g_{intra}(12)$ and interlayer $g_{inter}(12)$ distribution functions are related to the two-body densities through 
\begin{equation}
\left\{\begin{array}{ll} 
\displaystyle g_{intra}(12)&\displaystyle =1+h_{intra}(12)=\mbox{\Large{(}}\frac{4\pi}{\rho}\mbox{\Large{)}}^2 \rho_{intra}^{(2)}(s,\hat{\bm{\mu}}_1,\hat{\bm{\mu}}_2)\\
&\\
\displaystyle g_{inter}(12)&\displaystyle =1+h_{inter}(12) = \mbox{\Large{(}}\frac{4\pi}{\rho}\mbox{\Large{)}}^2\rho_{inter}^{(2)}(s,\hat{\bm{\mu}}_1,\hat{\bm{\mu}}_2)
\end{array}
\right. 
\end{equation}
In particular, the intralayer $g_{intra}^{000}(s)$ and interlayer $g_{inter}^{000}(s)$ center-to-center pair distribution functions are given by
\begin{equation}
\left\{\begin{array}{ll}
\displaystyle g_{intra}^{000}(s) &\displaystyle = \frac{1}{4\pi s\rho N_{0}}\Big<\sum_{i\in L_{1}}\sum_{j\in L_{1},j\neq i}\delta(s-\mid {\bm s}_{ij}\mid )+\sum_{i\in L_{2}}\sum_{j\in L_{2},j\neq i}\delta(s-\mid {\bm s}_{ij}\mid)\Big>\\
&\\
&\displaystyle  =\Big<g_{intra}(12)\Big>_{\hat{\bm{\mu}}_1\hat{\bm{\mu}}_2}\\
&\\
\displaystyle g_{inter}^{000}(s) &\displaystyle = \frac{1}{2\pi s\rho N_{0}}\Big<\sum_{i\in L_{1}}\sum_{j\in L_{2}}\delta(s-\mid {\bm s}_{ij}\mid)\Big>=\Big<g_{inter}(12)\Big>_{\hat{\bm{\mu}}_1\hat{\bm{\mu}}_2}
\end{array}
\right.
\end{equation}
where $\bm{s}_i$ is the in-plane position of particle $i$ according to the notations defined in Eq.(\ref{pot2}) and $< \cdot >_{\hat{\bm{\mu}}_1\hat{\bm{\mu}}_2}$ denotes  averaging over orientations of the dipole moments. The angular dependent pair correlation functions $h(12)$ have been expanded, as usual, on a basis set of rotational invariants $\tilde{\Phi}^{l_1 l_2 l}$ \cite{Blum:72,Gray:84}
\begin{equation}
\displaystyle h(12)=\sum_{l_1,l_2,l}h(l_1,l_2,l ; r)\tilde{\Phi}^{l_1 l_2 l}(\hat{\bm{\mu}}_1,\hat{\bm{\mu}}_2,\hat{\bm{r}})
\end{equation}
where the $\tilde{\Phi}^{l_1 l_2 l}$ are related to the standard rotational invariants $ \Phi^{l_1 l_2 l}$ in an expansion on spherical harmonics by  (see e.g. \cite{Patey:77})
\begin{equation}
\displaystyle \tilde{\Phi}^{l_1 l_2 l}=\frac{1}{l !}\left( \begin{array}{lll} l_1 & l_2 & l\\ 0 & 0 & 0 \end{array}\right) \Phi^{l_1 l_2 l}.
\end{equation} 
The most significant projections of the intralayer $h_{intra}(12)$ and interlayer $h_{inter}(12)$ correlation functions calculated in this work are those onto $\tilde{\Phi}^{110}$ , $\tilde{\Phi}^{112}$ and $\tilde{\Phi}^{220}$. The correponding expressions are summarized in Table \ref{tb:coef}.\\

{\centering
\subsection{Order parameter}
}
Possible orientational (nematic) order in a layer can be established from the non-vanishing of  the second-rank order parameter $P_2$ calculated as the average value of the largest eigenvalue of the matrix
\cite{Eppeng:84} 
\begin{equation}
\displaystyle Q_{\alpha \beta} = \frac{1}{N_0} \sum_i^{N_0} \frac{1}{2}(3 {\hat{\mu}}_{\alpha}^i {\hat{\mu}}_{\beta}^i-{\delta}_{\alpha \beta}),
\end{equation}
where ${{\hat{\mu}}_{\alpha}}^i$ is the $\alpha$ component of the unit vector ${\hat{\bm{\mu}}}_i$. One can note that the projection $h^{220}$ obeys the asymptotic relationship  
\begin{equation}
h^{220}(s)  \sim 5 P_2^2,  \ \ s \to \infty
\label{h220limit}
\end{equation}
As will be shown below no global nematic order occurs in the systems for $\rho < 0.7$. 

\begin{figure}[htbp]
  \begin{center}
    \mbox{
      \subfigure{\scalebox{0.3}{\includegraphics[width=15.1in]{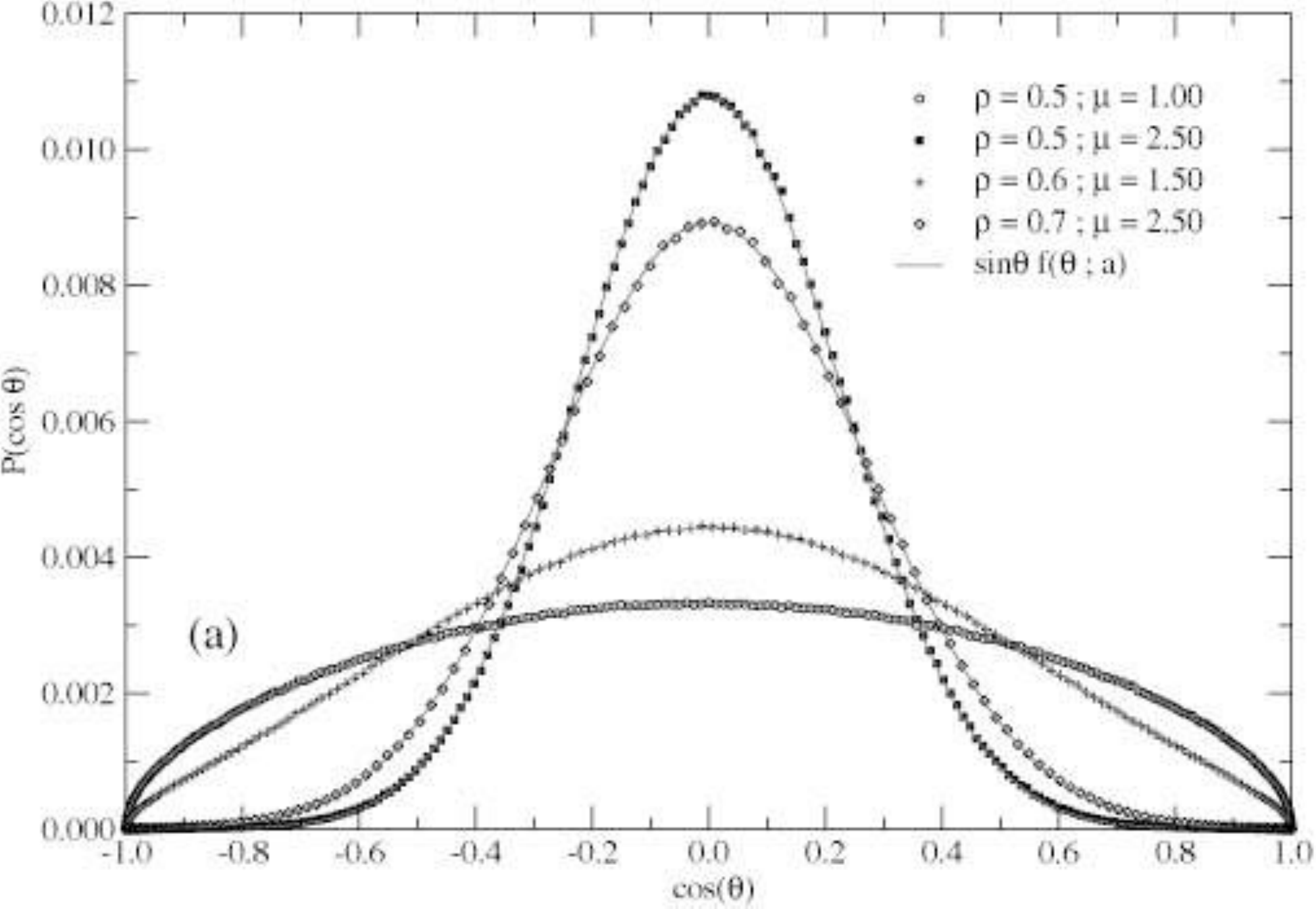}}}
    }
    \hspace{2.in}
    \mbox{
      \subfigure{\scalebox{0.3}{\includegraphics[width=15.in]{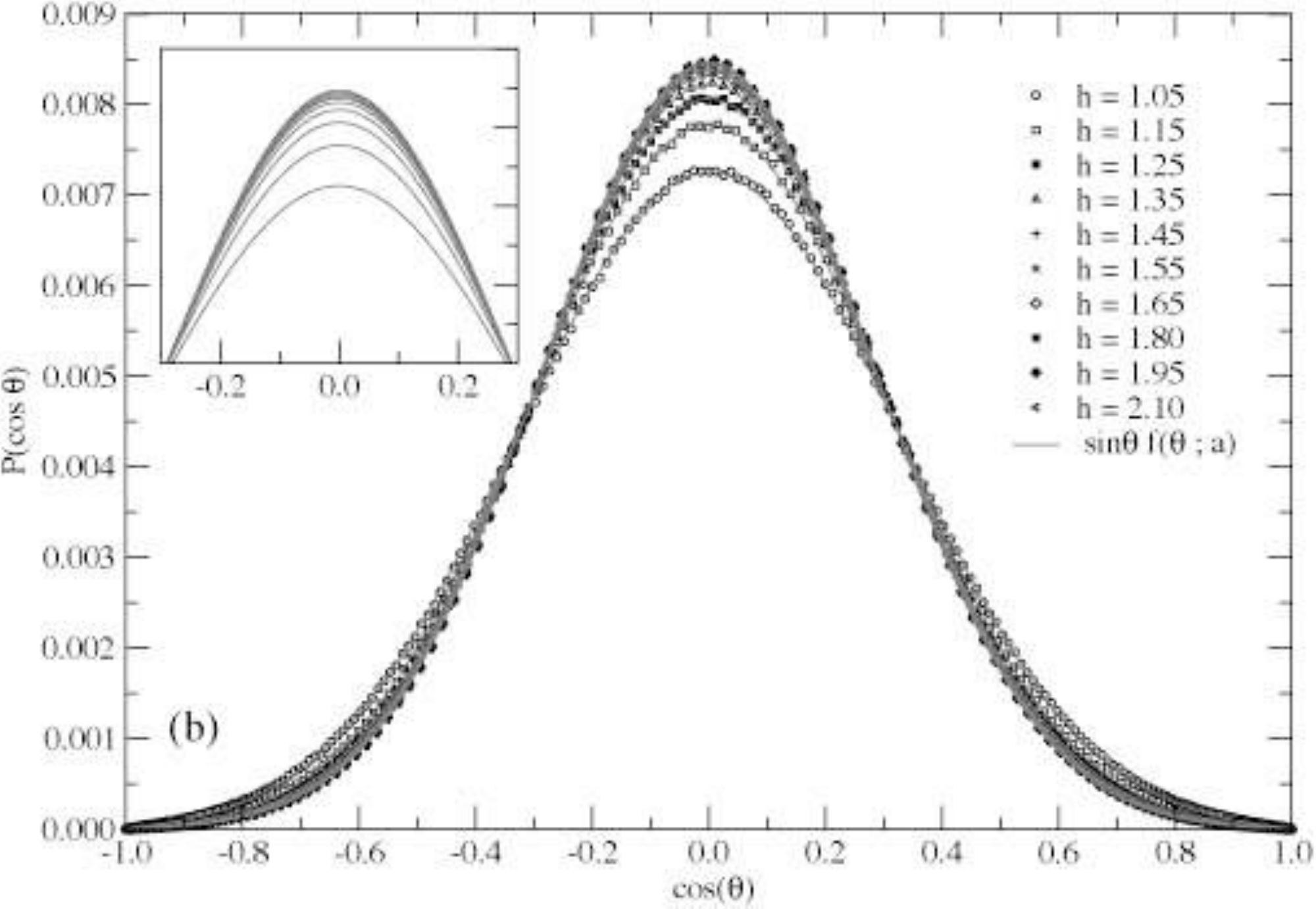}}}
    }
    \caption{Orientational distribution functions of dipolar moments in a bilayer of dipolar hard spheres. Symbols denote MC data and solid lines are fits using Eq.(\ref{fit1}). (a) Results at selected values of $\rho$ and $\mu$ at $h=1.05$. (b) Variation with layer separation $h$ for $\rho = 0.7$ and $\mu=2.00$. }
    \label{fig.1}
  \end{center}
\end{figure}

{\centering
\section{RESULTS }\label{Results}
}
{\centering
\subsection{One-body orientational distribution function}
}
One-body distribution functions $f({\hat{\bm{\mu}}}) = f(\cos(\theta))$, with  polar angle $\theta$  defined by  $\cos \theta = \hat{\bm{\mu}}\cdot\hat{\bm{e}}_z$,   obtained from MC simulation at various thermodynamic states are shown in Fig. \ref{fig.1}(a). It is seen that for all states an excellent fit to the MC data is obtained with the one  parameter function 
\begin{equation}
\label{fit1}
\displaystyle f(\cos \theta ; a) = f_0 \exp(-a\cos^2\theta)
\end{equation}
with normalization constant 
\begin{equation}
\label{fit1a}
\displaystyle f_0=\sqrt{\frac{a}{\pi}}\frac{1}{\mbox{ erf}(\sqrt{a})}
\end{equation}
Values of $a$ obtained  by fitting the MC histograms $P(cos \theta)$, normalized to one, are given  in Tables \ref{tb:varh}-\ref{tb:Pfixh}. The results for the orientational distribution functions of the  bilayer system are quite similar to those obtained  earlier for monolayers \cite{Weis:02a}. As $\mu$ increases the dipole moments tilt more and more into the layer plane ($\cos \theta \sim 0$). The interaction between the two layers induces, though, a  slight effect, in comparison to the monolayer system, as seen in Fig. \ref{fig.1}(b) showing the variation of the orientational distributions with interlayer separation $h$ for $\rho=0.7$ and $\mu=2.00$. As the separation between the layers decreases, the coupling between layers increases which entails a slight tendency of the dipoles to orient perpendicularly to the plane. As a consequence the distributions are slightly broadened (the value of $a$ decreases).

\begin{figure}[htbp]
  \begin{center}
    \centerline{\includegraphics[width=7.in]{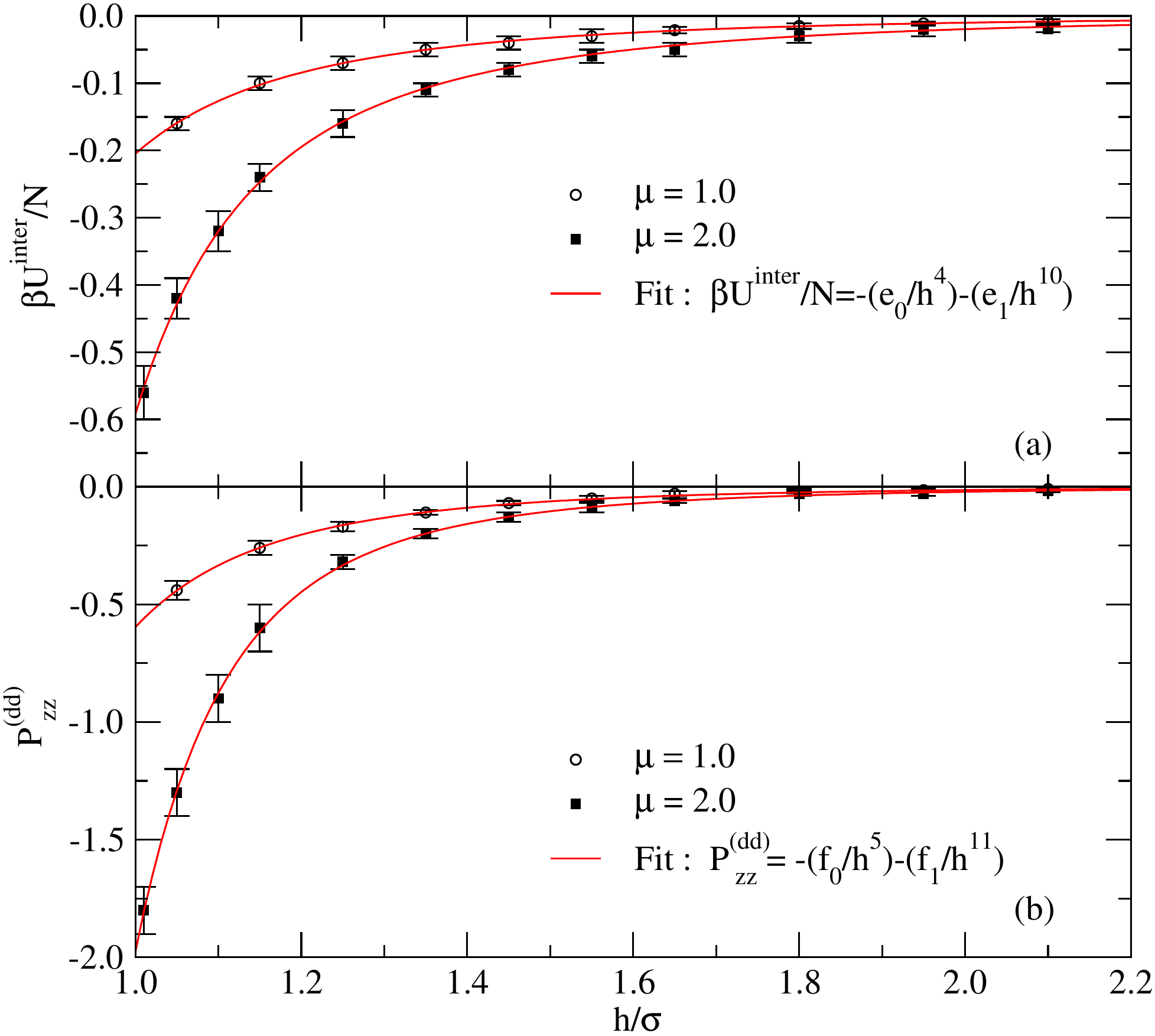}}
    \caption{Average energies (a) and normal pressures (b) as a function of $h$ for $\rho=0.7$ and $\mu=1$ and 2. The symbols denote MC data and the lines are fits to the data using Eqs.(\ref{fit2}) and (\ref{fit3}), respectively. The fitting parameters for $\mu$=1 are $e_0= 0.16\pm 0.01$, $e_1= 0.045\pm 0.002$ and $f_0= 0.46\pm 0.01$, $f_1= 0.13 \pm 0.01 $. For $\mu$=2 they are  $e_0=0.31\pm 0.01$, $e_1= 0,28\pm 0.01$ and $f_0= 0.68\pm 0.03$, $f_1= 1.3\pm 0.1$.}
    \label{fig.2}
  \end{center}
\end{figure}

{\centering
\subsection{Energy}
}
The variation of the intralayer $\beta U^{intra} / N$ and interlayer $\beta U^{inter} / N$ energies as a function of layer separation are summarized in Table \ref{tb:varh}  for the density $\rho=0.7$ and the two dipole moments $\mu$ = 1 and 2. The intralayer energy is seen to be by far the dominant contribution and is nearly independent  of $h$ especially at the largest dipole moments where in-plane orientation of the dipole moments is prevalent. The interlayer energy is much smaller and decreases rapidly with layer separation vanishing at $h \approx 2$. The total energy remains practically constant when $h$ varies from 1.05 to 2.0.\\
Attard and Mitchell have applied a second order perturbation theory on a bilayer of orientable dipoles \cite{Attard:87,Attard:88} and found that the interaction free energy between the surfaces decays as the fourth power of $h$ at large separation. An analysis of our MC data, for $h\gtrsim 1.6$, agrees with the behavior obtained in the computations done by Attard and Mitchell ; more precisely, the variation of the interlayer energy with $h$, for $\rho=0.7$ and $\mu$=1 and 2, can be quite well represented by
\begin{equation}
\label{fit2}
\displaystyle \frac{\beta U^{inter}}{N} =-\frac{e_0}{h^4}-\frac{e_1}{h^{10}}
\end{equation}
where $e_0$ and $e_1$ are obtained by a fit to the simulation results (see Fig.\ref{fig.2}(a)).\\
Table \ref{tb:Ufixh} summarizes energy values obtained at fixed layer separation $h=1.05$ for various dipole moments in the density range $\rho=0.3-0.7$. For all densities considered the intralayer energy decreases with $\mu$ and saturates near $\mu \approx 2.5$. The variation with density diminishes when the  dipole moment is increased. The interlayer energy is much smaller than the intralayer contribution presenting, at all densities, a shallow minimum in the range $\mu \approx 1.75-2.0 $ where appreciable chaining of the particles sets in.

\begin{figure}[htbp]
  \begin{center}
	\centerline{\includegraphics[width=5.in]{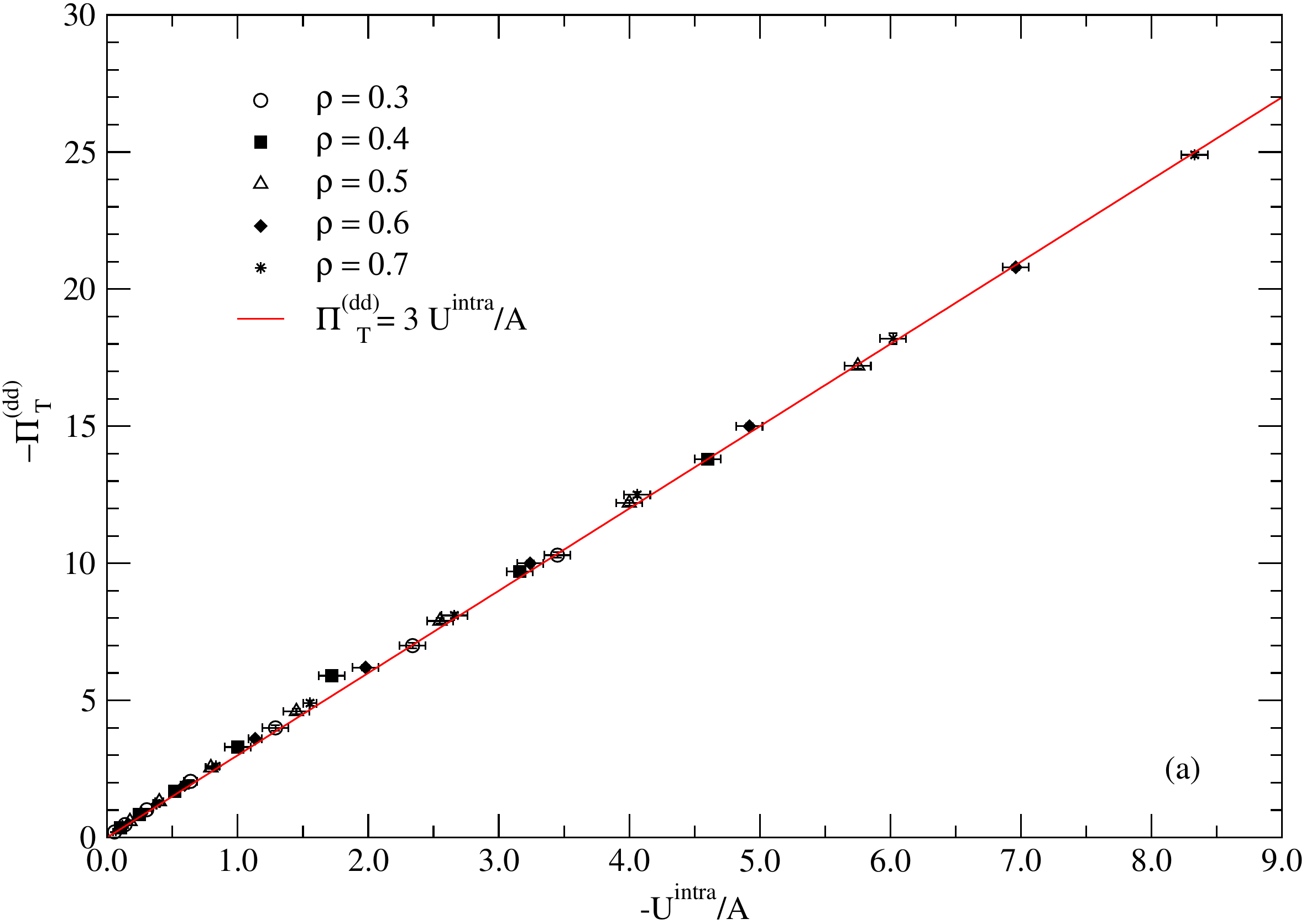}}
	\centerline{\includegraphics[width=5.in]{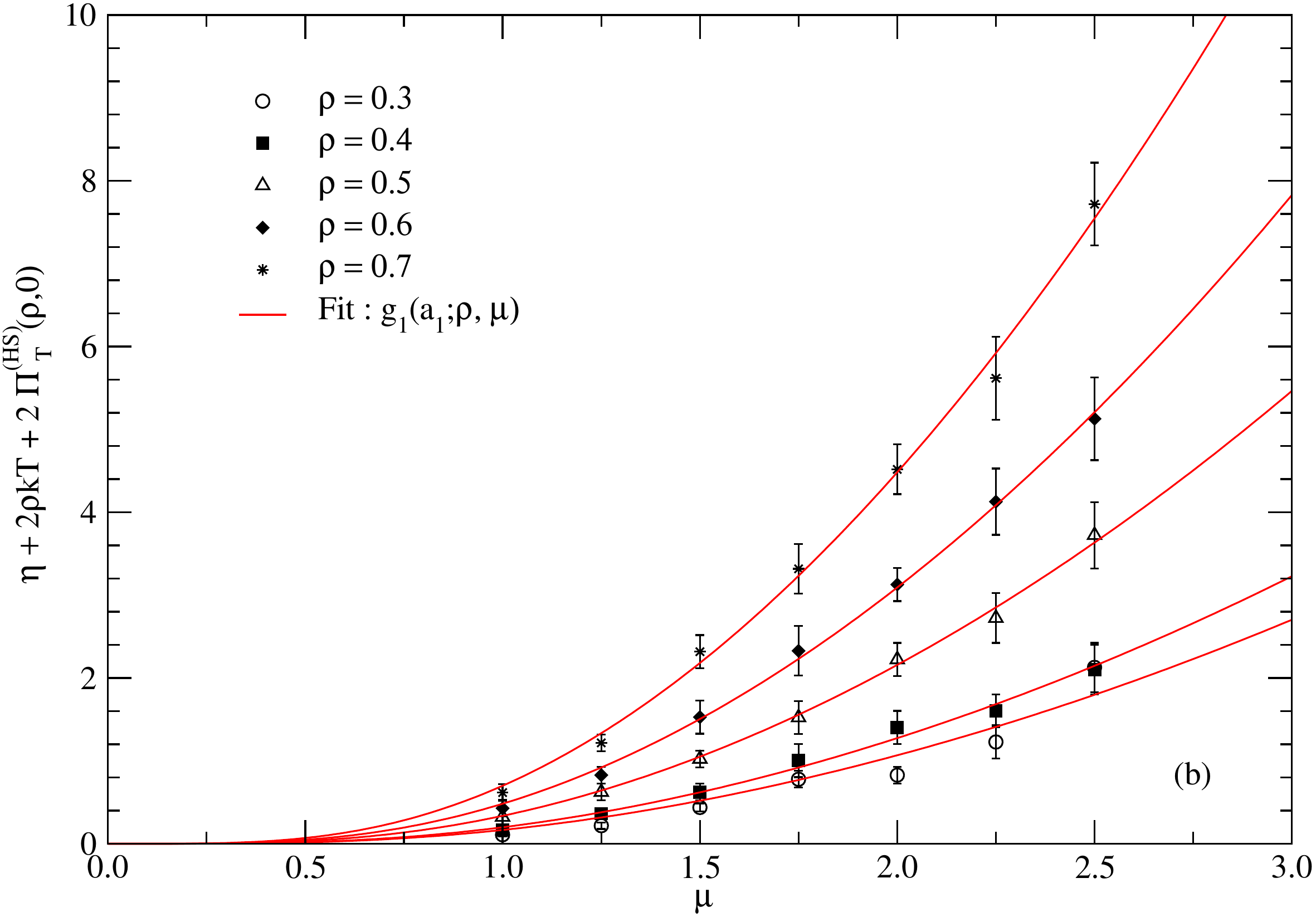}}
    \caption{(a) Lateral pressure as a function of the intralayer energy
    per unit area. Symbols are data from Tables III and IV for densities
    $\rho=0.3-0.7$, dipole strengths $\mu=1.0-2.5$ and $h=1.05$ ; the
    straight line is given by Eq.(\ref{eos1}). (b) Surface stress as
    function of dipole strength for $\rho=0.3-0.7$ and
    $h=1.05$. Symbols are data from Table IV and lines are given by
    Eq.(\ref{fit4}) with $g_1(a_1 ;\rho,\mu)=a_1\rho^2\mu^4/(1+\mu^2)$
    ($a_1\sim 2.7$) and $\Pi_T^{(HS)}(\rho,0)$ given by the equation of
    state of hard disks ref.\ \cite{Santos:95}.}

    \label{fig.3}
  \end{center}
\end{figure}

{\centering
\subsection{Pressure and surface stress}
}
Similar to the interlayer energy, the normal pressure at constant $\mu$ and $\rho$ is quite well represented, as a function of $h$, by
\begin{equation}
\label{fit3}
\displaystyle P_{zz}=-\frac{f_0}{h^5}-\frac{f_1}{h^{11}}
\end{equation}
However, as for a thermodynamical variable $X$ generally  
\begin{center}
$\displaystyle \Big <\frac{\partial X}{\partial h} \Big > \neq \frac{\partial <X>}{\partial h}$, 
\end{center}
the fitting parameters $f_0$ and $f_1$ for the pressure do not relate directly to those for the energy. Nevertheless, the functional form of Eq.(\ref{fit3}) obtained as the derivative of Eq.(\ref{fit2}) provides quite good agreement between simulation results and Eq.(\ref{fit3}) (see Fig \ref{fig.2}(b)).\\
As seen in Table \ref{tb:varh}, the surface stress, for $\rho=0.7$, is fairly independent of  $h$ for $\mu=1$ and 2. For $\mu=1$, all the thermodynamic quantities, $\Pi_{T}^{(dd)}$ and $\Pi_{T}^{(HS)}$, that contribute to $\tilde{\eta}$ through Eq.(\ref{BB15}) and (\ref{press2}), are nearly constant. For $\mu=2$, $\tilde{\eta}$ appears also to be insensitive to $h$, but a small counterbalance between $\Pi_{T}^{(dd)}$ and $\Pi_{T}^{(HS)}$ is observed as $h$ increases from 1.01 to 1.15. As apparent from the one body orientational distribution functions, for $h$ between 1.01 and 1.15 and  $\mu=2$, the dipoles are on average less parallel to the layers than would be the case for larger $h$ values. Thus, the attraction between particles in the same layer is slightly decreased in comparison to a monolayer ; this increases $\Pi_{T}^{(dd)}$ and reduces $\Pi_{T}^{(HS)}$, since less contact between particles are observed in $g_{intra}^{000}(\sigma)$. One should note, though, that this effect is quite small (see Table \ref{tb:varh}).\\
The values of $\tilde{\eta}$, for $\mu=1$, $\rho=0.7$ and $h>2.00$, given in table \ref{tb:varh}, agree with the results obtained for the 2D pressure of the monolayer (see Tables I and II in ref.\cite{Lomba:00} - as outlined in subsection 3.2, the value of $\Pi_T$ obtained from $\tilde{\eta}$ is twice the value of the pressure found in ref.\cite{Lomba:00}).\\ 
As shown previously, the 2D pressure of a monolayer of DHS may be related to the internal energy of the monolayer (see Eq.(21) in ref.\cite{Lomba:00}). For the bilayer, we obtain almost exactly the same result, except for a factor 2 discussed before in subsection III.B. In Fig.\ref{fig.3}(a), we have represented $-\Pi_{T}^{(dd)}$ as a function of $-U^{intra}/A$ ; it appears that the dipolar contribution to the lateral pressure of the bilayer is very well represented by
\begin{equation}
\label{eos1}
\Pi_{T}^{(dd)}=3\rho kT\mbox{ }\frac{\beta U^{intra}}{N}=3\frac{U^{intra}}{A}.
\end{equation}     
Thus, for $\rho\leq 0.7$ and $\mu\leq 2.5$, the equation of state is given by an equation similar to Eq.(21) of ref.\cite{Lomba:00} as
\begin{equation}
\label{eos2}
\frac{\Pi_{T}}{2\rho kT}=1+\frac{\Pi_{T}^{(HS)}}{\rho kT}+\frac{3}{2}\frac{\beta U^{intra}}{N}=-\frac{\tilde{\eta}}{2\rho kT}.
\end{equation}
The variation of  $\tilde{\eta}$ with dipole moment is shown in Fig.\
\ref{fig.3}(b) for $h=1.05$ and various densities. $\tilde{\eta}$ can be
approximated empirically  by relations as 
\begin{equation}
\label{fit4}
\displaystyle \tilde{\eta}(\rho,\mu) =-2\rho kT - 2\Pi_T^{(HS)}(\rho,0)+g(a_1 ;\rho,\mu)
\end{equation}
where $g(a_1;\rho,\mu)$ is a function of the fitting parameter $a_1$ and $\Pi_T^{(HS)}(\rho,0)$ obtained from the equation of state of hard disks (see, for instance, ref.\cite{Santos:95}). Several functional forms for $g$, as for instance, $g_1 (a_1 ;\rho,\mu)=a_1\rho^2\mu^4/(1+\mu^2)$, with $a_1\sim 2.7$, or $g_2(a_2 ;\rho,\mu)=a_2\rho^2\mu^{5/2}$, with $a_2\sim 1.6$ were found to reproduce quite accurately the numerical results given in Table IV.

\begin{figure}[htbp]
  \begin{center}
    \centerline{\includegraphics[width=7.in]{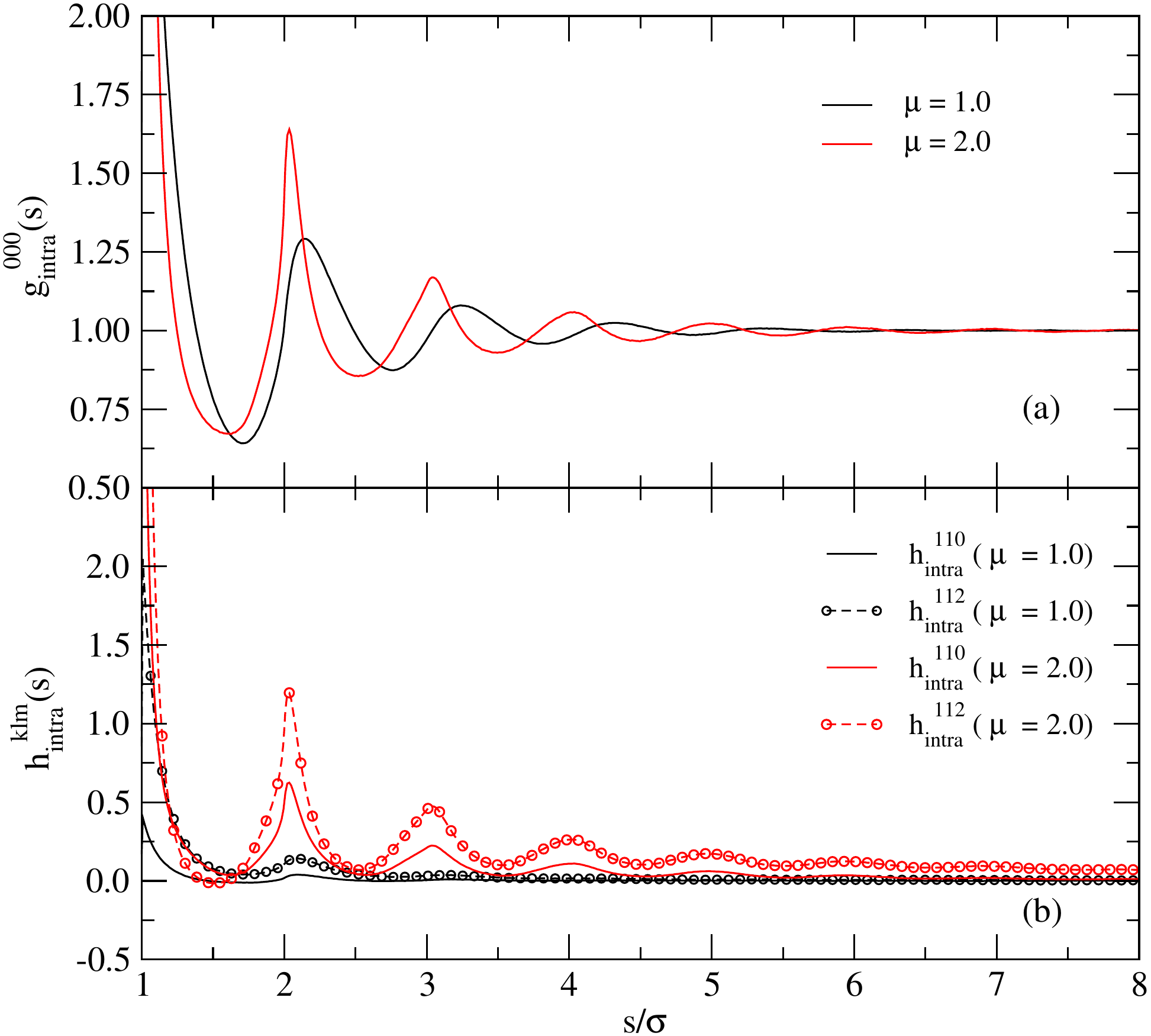}}
    \caption{Intralayer angle averaged pair distribution function (a) $g_{intra}^{000}(s)$ and angular projections (b) $h_{intra}^{klm}(s)$ for a bilayer of dipolar hard spheres at  $\rho=0.7$, $h=1.05$ for $\mu=1.0$ (black) and $\mu=2.0$ (red -grey hue).  }
    \label{fig.4}
  \end{center}
\end{figure}

\begin{figure}[htbp]
  \begin{center}
    \centerline{\includegraphics[width=7.in]{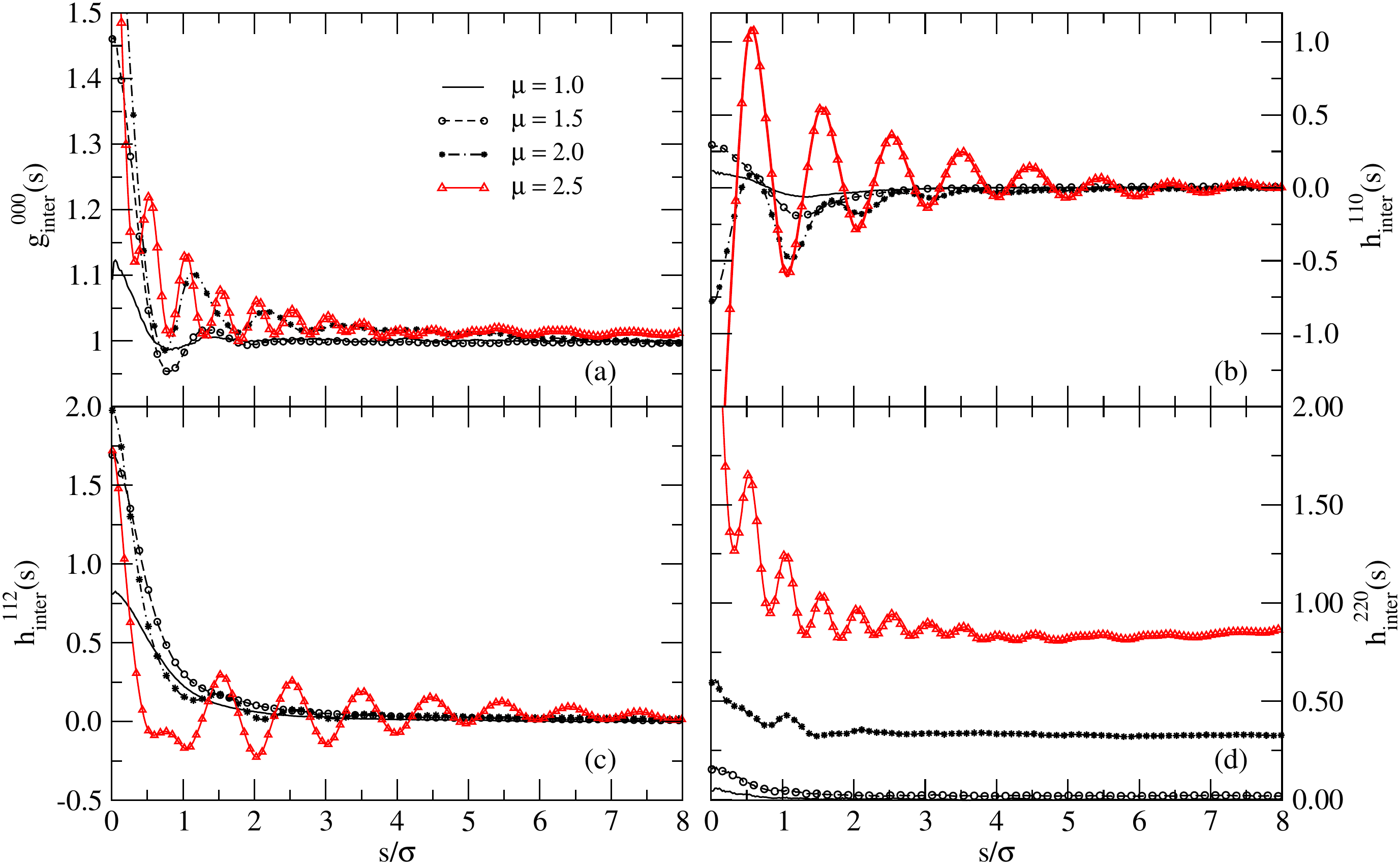}}
    \caption{Interlayer angle averaged pair distribution function $g_{inter}^{000}(s)$ and angular projections $h_{inter}^{klm}(s)$ of the pair distribution functions $g_{inter}(12)$ for the DHS bilayer at $\rho=0.3$ and $h=1.05$ for several values of $\mu$. (a) $g_{inter}^{000}(s)$ ; (b) $h_{inter}^{110}(s)$ ; (c)  $h_{inter}^{112}(s)$, (d) $h_{inter}^{220}(s)$.}
    \label{fig.5}
  \end{center}
\end{figure}

\begin{figure}[htbp]
  \begin{center}
    \centerline{\includegraphics[width=7.in]{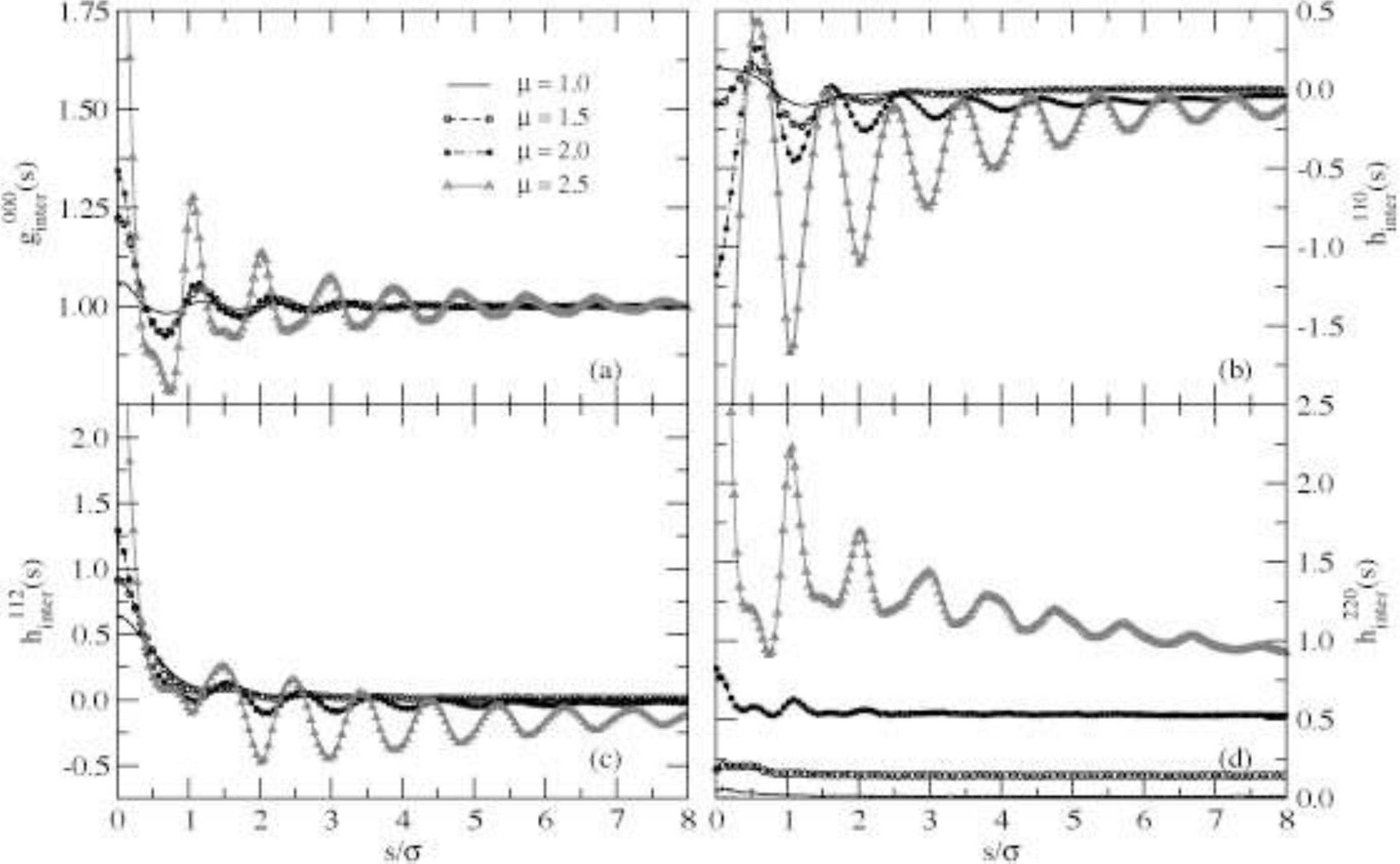}}
    \caption{Same as Fig. \ref{fig.5} but for $\rho=0.7$.}
    \label{fig.6}
  \end{center}
\end{figure}

\begin{figure}[htbp]
  \begin{center}
    \mbox{
       \subfigure[]{\scalebox{0.3}{\includegraphics[width=9.in]{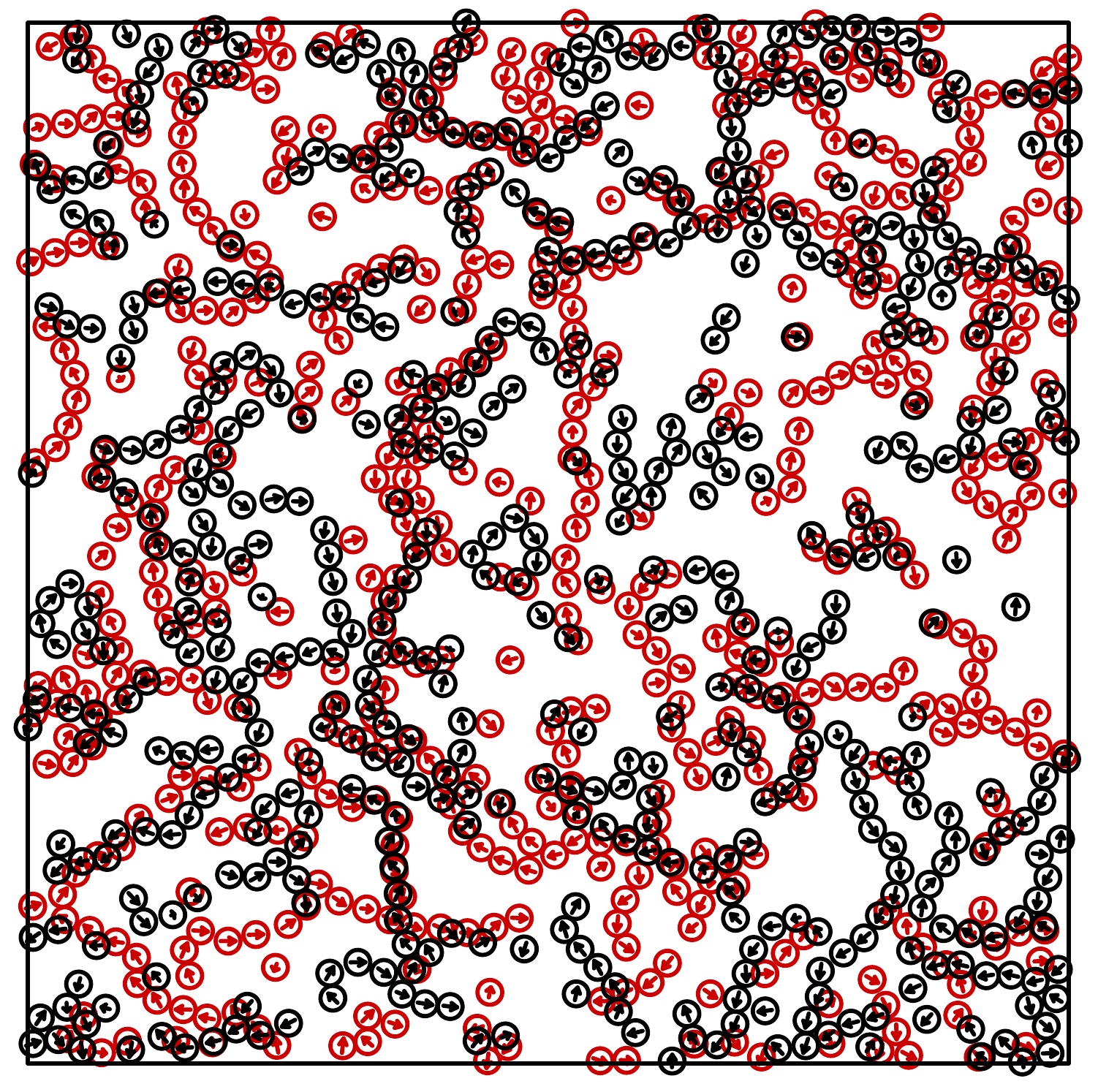}}} \quad
       \subfigure[]{\scalebox{0.3}{\includegraphics[width=9.in]{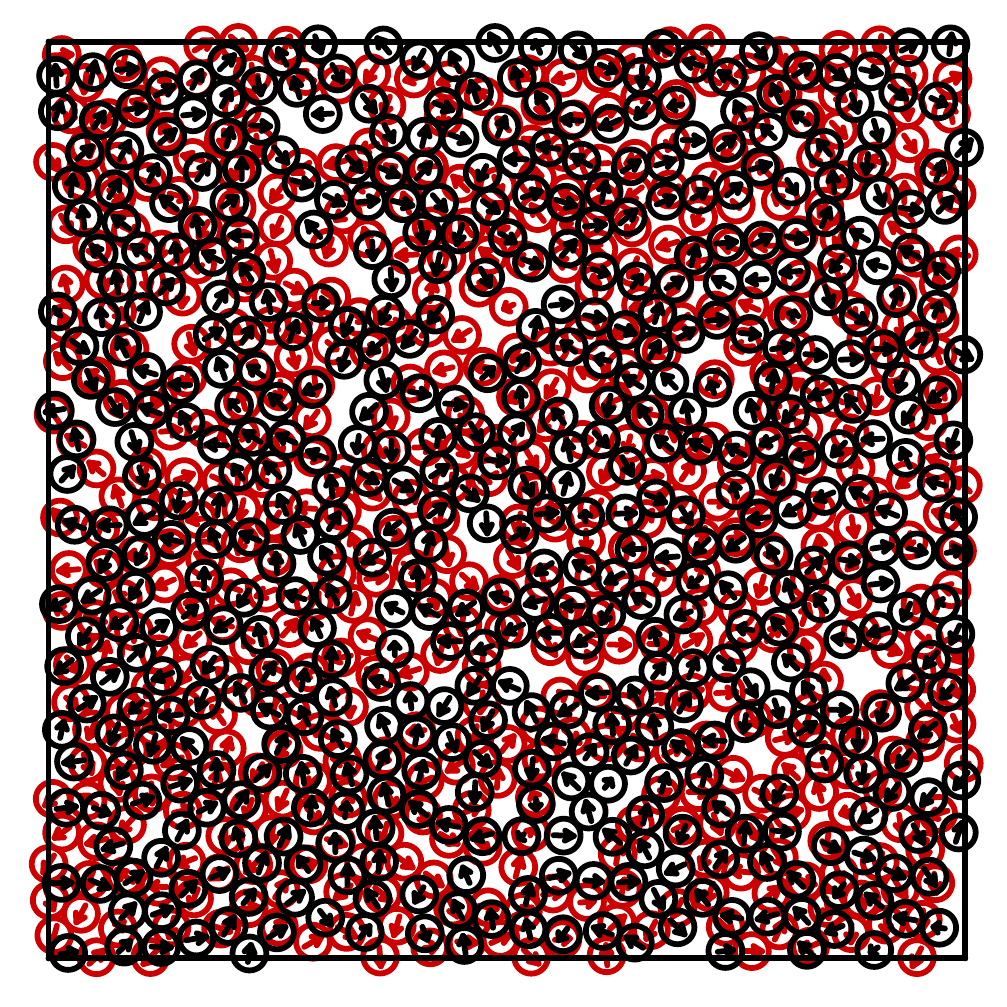}}}
    }\\
    \vspace{1cm}
    \hspace{.5cm}
    \mbox{
      \subfigure[]{\scalebox{0.3}{\includegraphics[width=9.in]{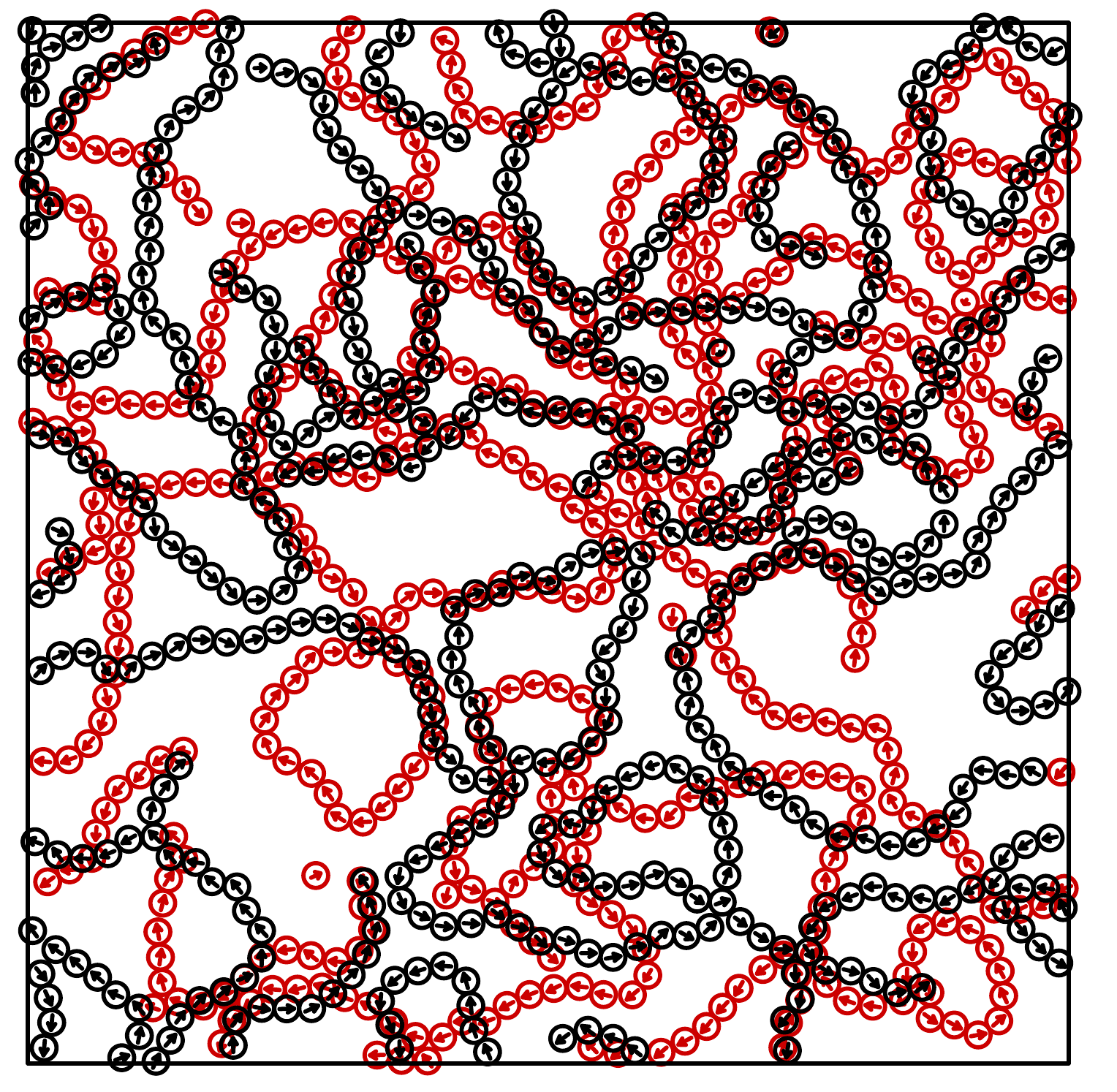}}} \quad
      \subfigure[]{\scalebox{0.3}{\includegraphics[width=9.in]{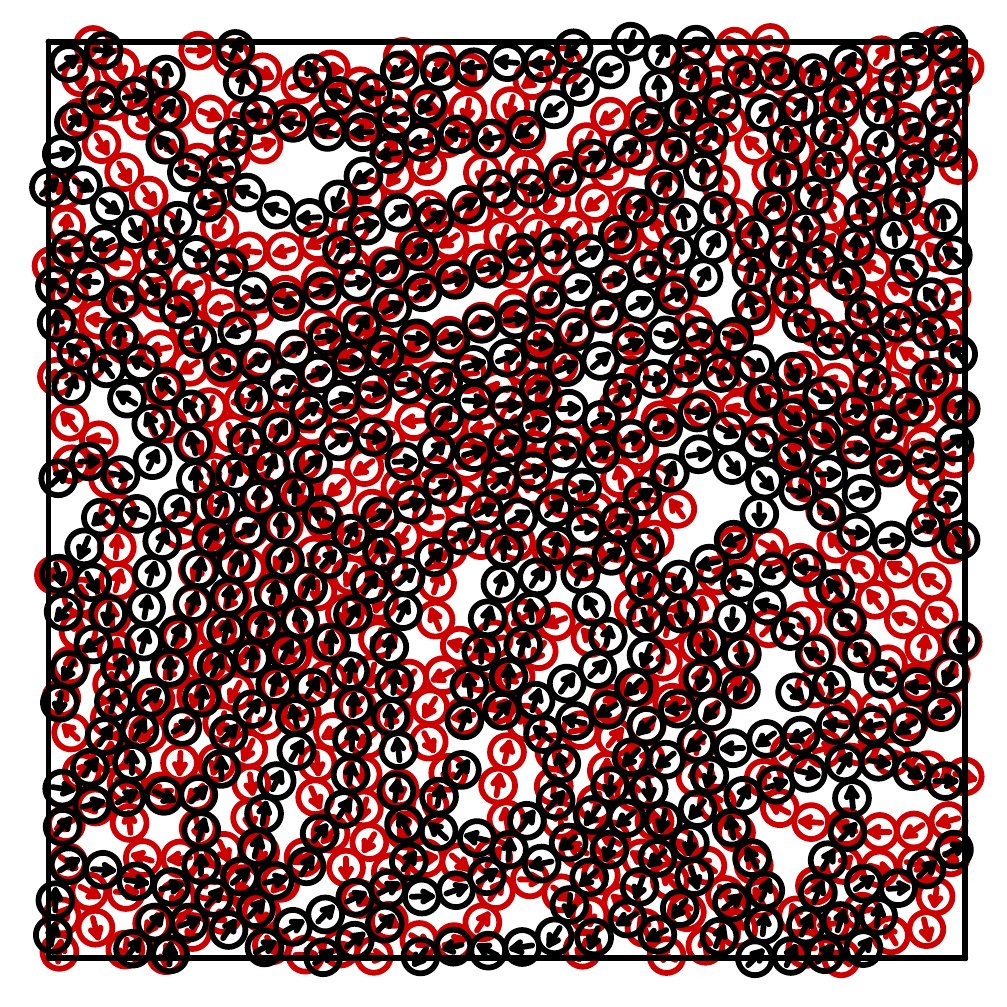}}}
    }
    \caption{Snapshots of  bilayer configurations of particles at
    $\mu=2.0$ ((a),(b)) and $\mu=2.50$ ((c),(d)) for  $h=1.05$ ;
    snapshots (a) and (c) are for
    $\rho=0.3$ ($N=1058$); snapshots (b) and (d) for $\rho=0.7$
    ($N=1024$). Particles in 
    different layers are represented by  different colours. The HS
    cores are represented by circles of diameter $\sigma=1$ and the
    directions of dipole moments by arrows.}
    \label{fig.7}
  \end{center}
\end{figure}

\begin{figure}[htbp]
  \begin{center}
    \mbox{
      {\scalebox{0.3}{\includegraphics[width=9.in]{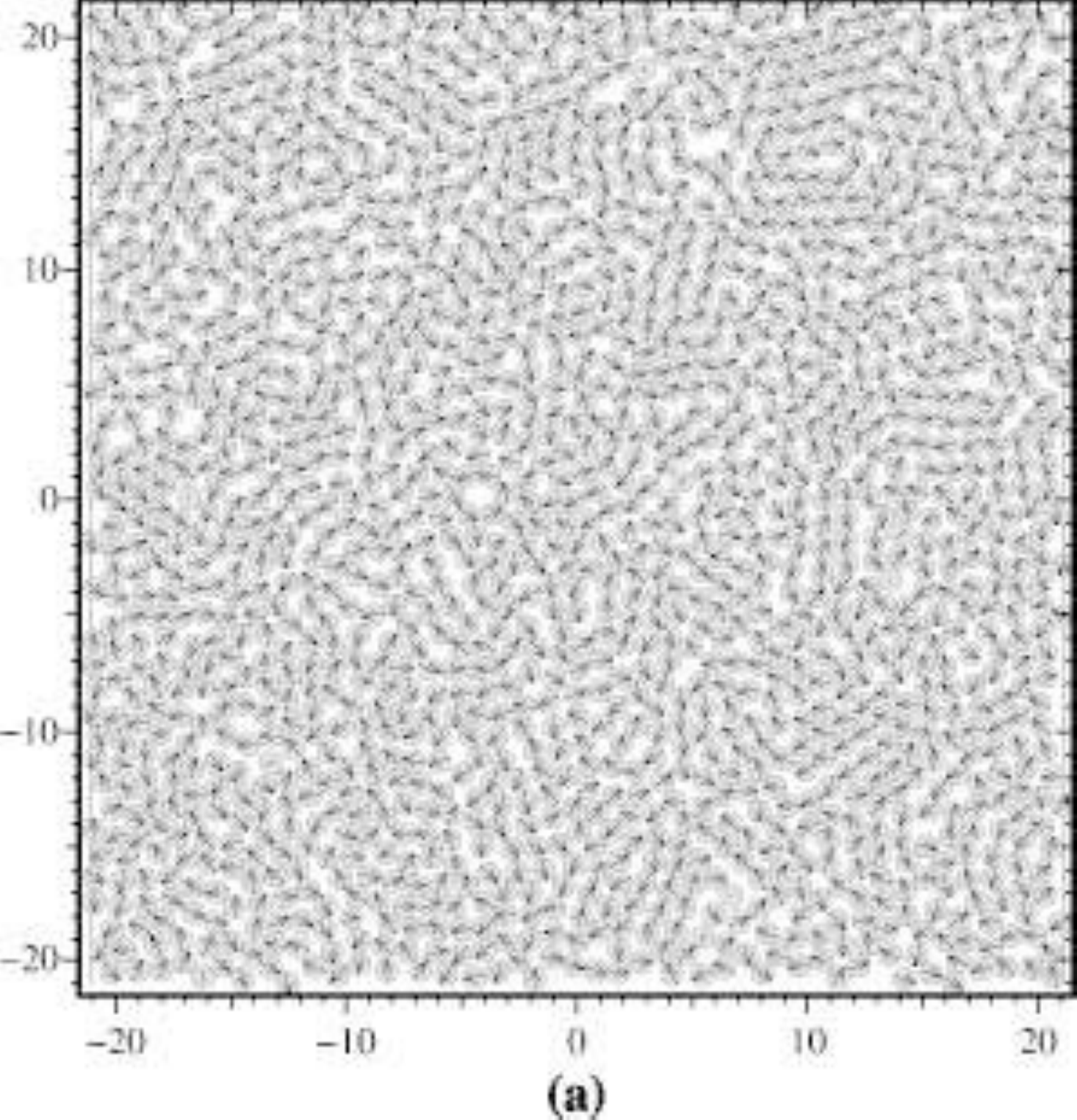}}} \quad
      {\scalebox{0.3}{\includegraphics[width=9.in]{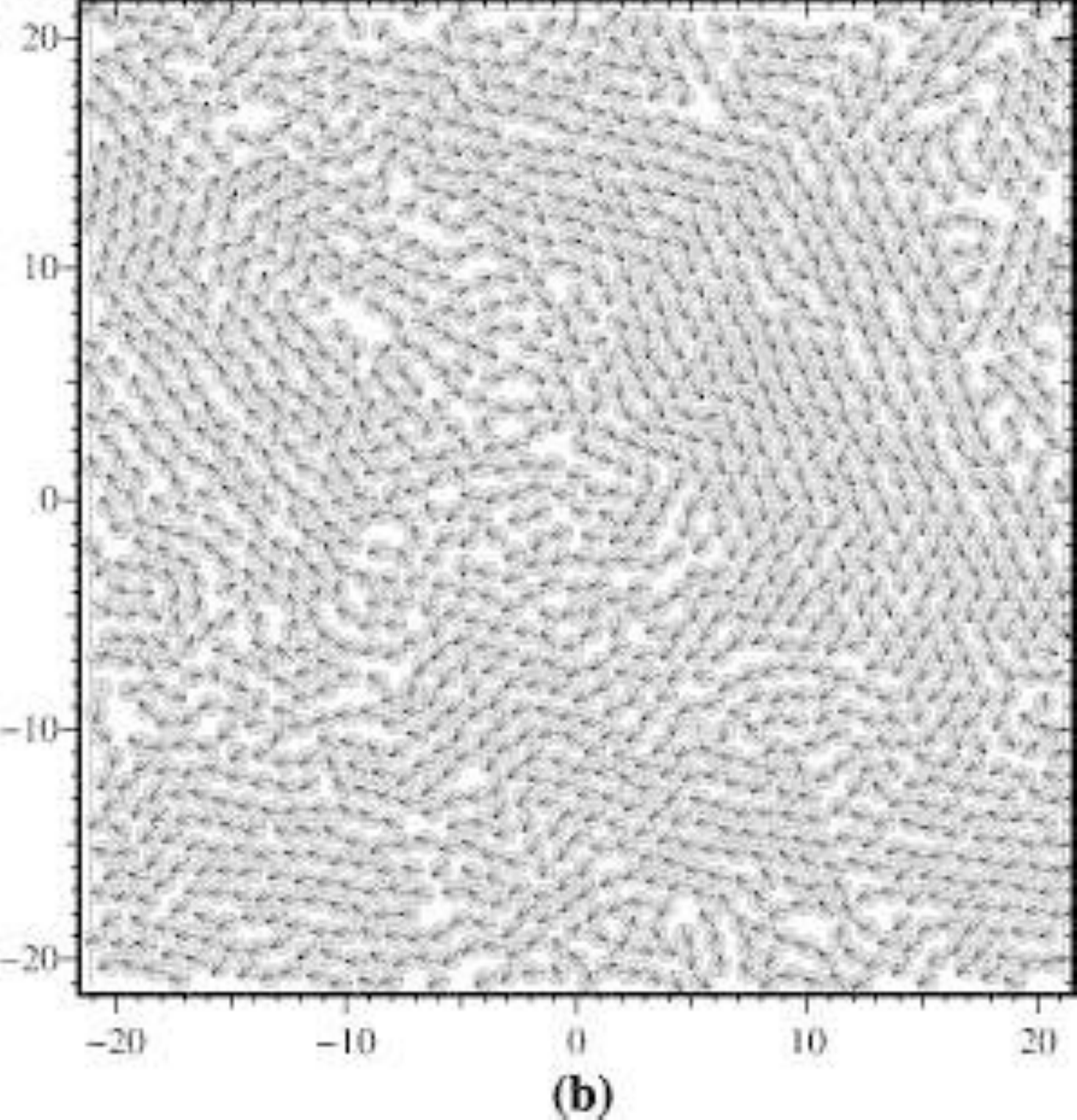}}}
    }\\
    \vspace{1cm}
    \hspace{.5cm}
    \mbox{
      {\scalebox{0.3}{\includegraphics[width=9.in]{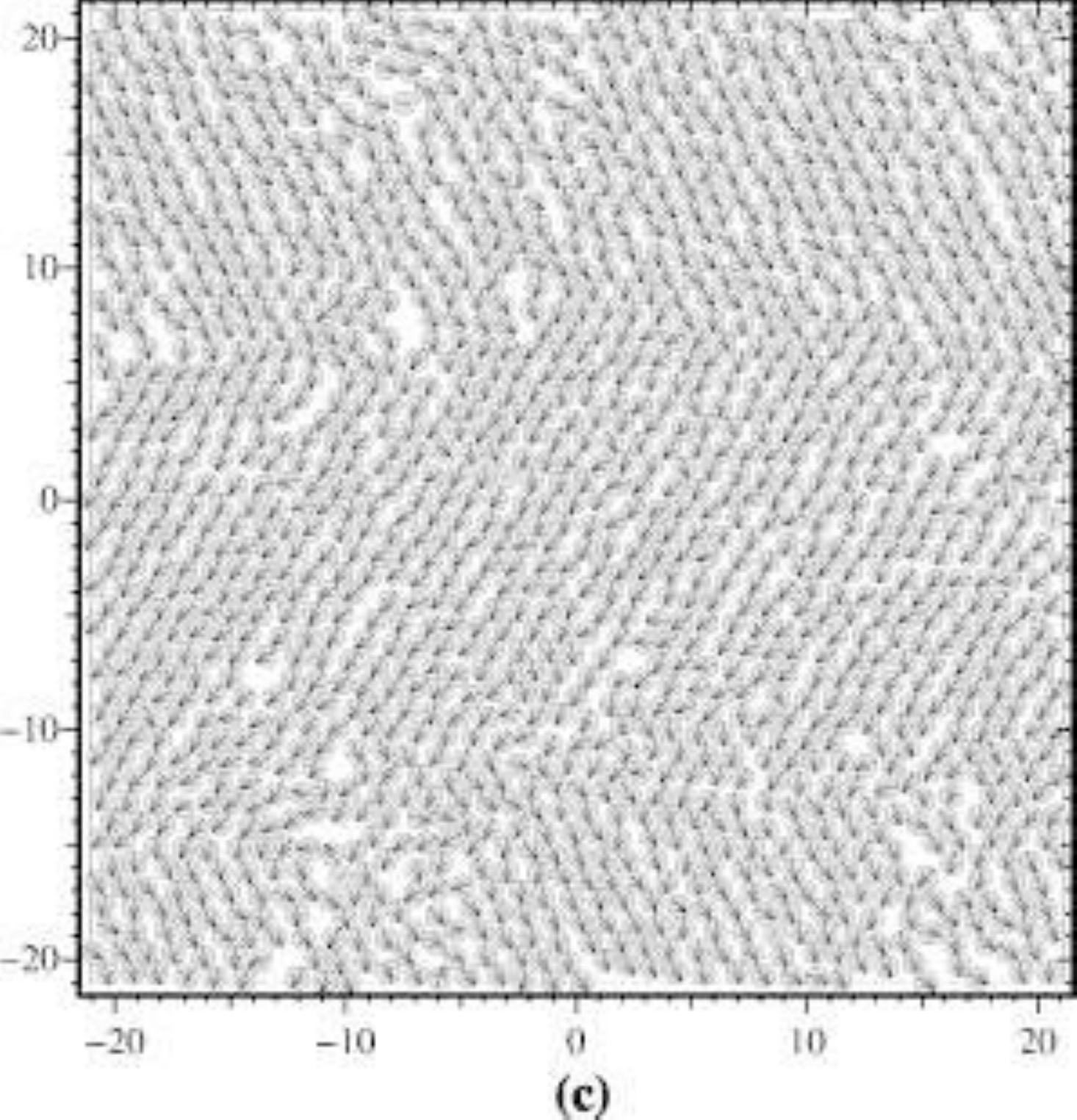}}} \quad
     {\scalebox{0.3}{\includegraphics[width=9.in]{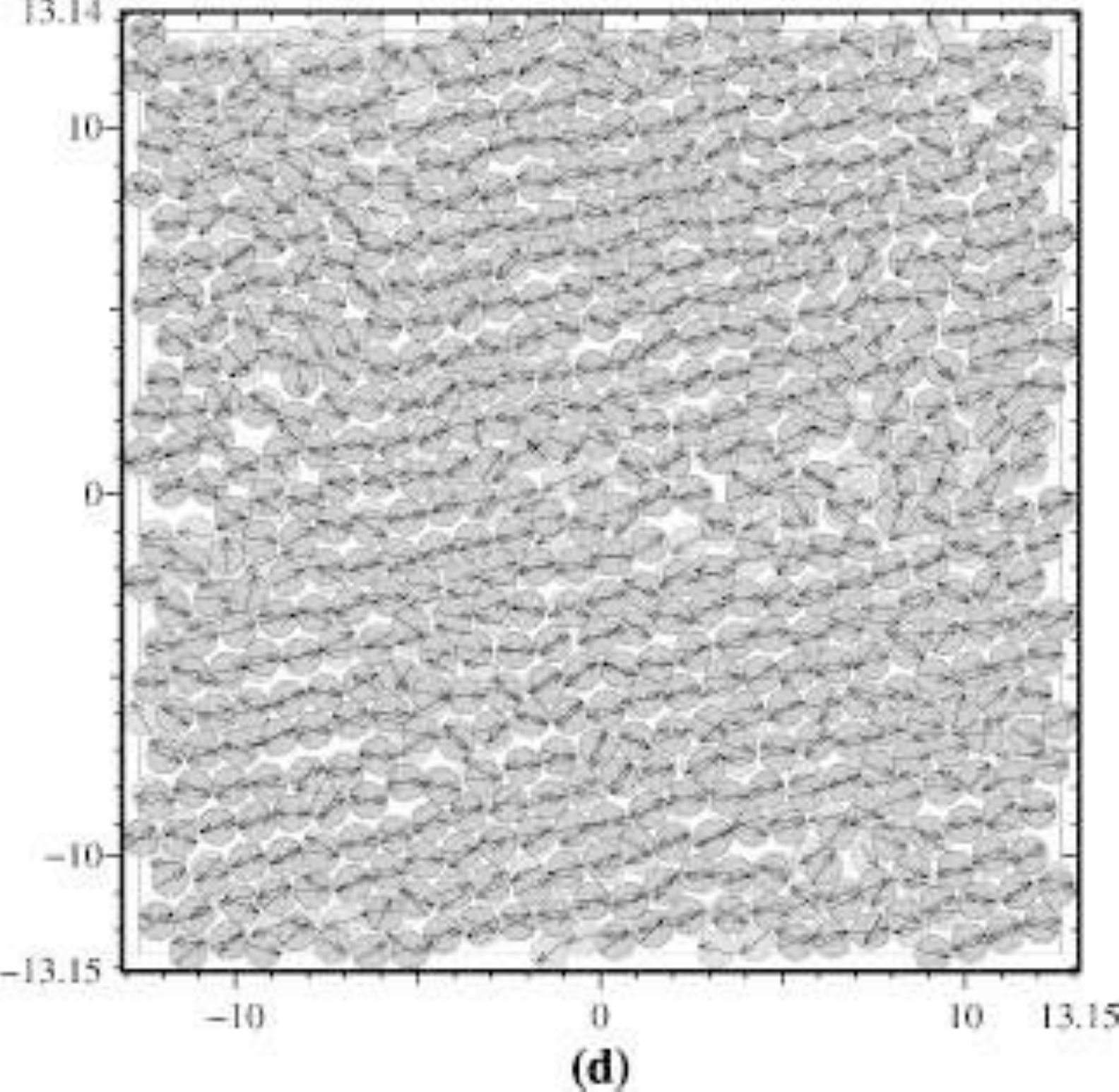}}}
    }
    \caption{Bilayer configurations of the $2 \times 1600$ particle system  at
    $\rho=0.9$,  $\mu=2$ and  $h=1.05$  at different intervals of the MC
    simulation; (a) snapshot after 500 cycles, (b) $0.26 \times 10^6$
    cycles, (c) $1.75 \times 10^6$ cycles, (d) result for $2 \times 576$
    particles 
    after  $2.6 \times 10^6$ cycles. For clarity only the particle
    arrangements in  one layer are shown in  (a)-(c). The
    arrows denote the projections of the dipole moments on the layer
    plane. Thus dipoles perpendicular to the layer appear as dots. }

    \label{fig.8}
  \end{center}
\end{figure}

\begin{figure}[htbp]
  \begin{center}
    \mbox{
      {\scalebox{0.3}{\includegraphics[width=9.in]{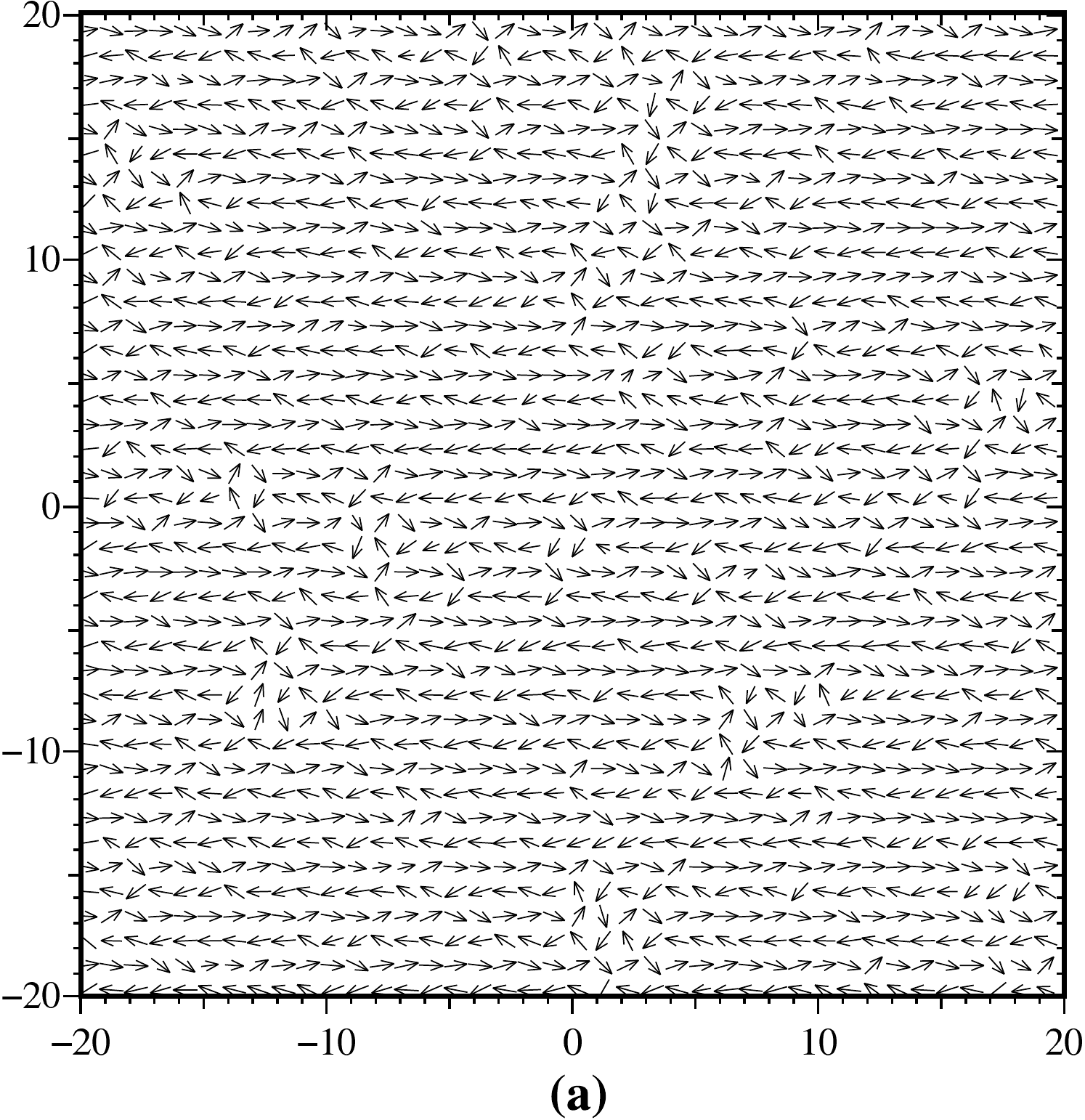}}} \quad
      {\scalebox{0.3}{\includegraphics[width=9.in]{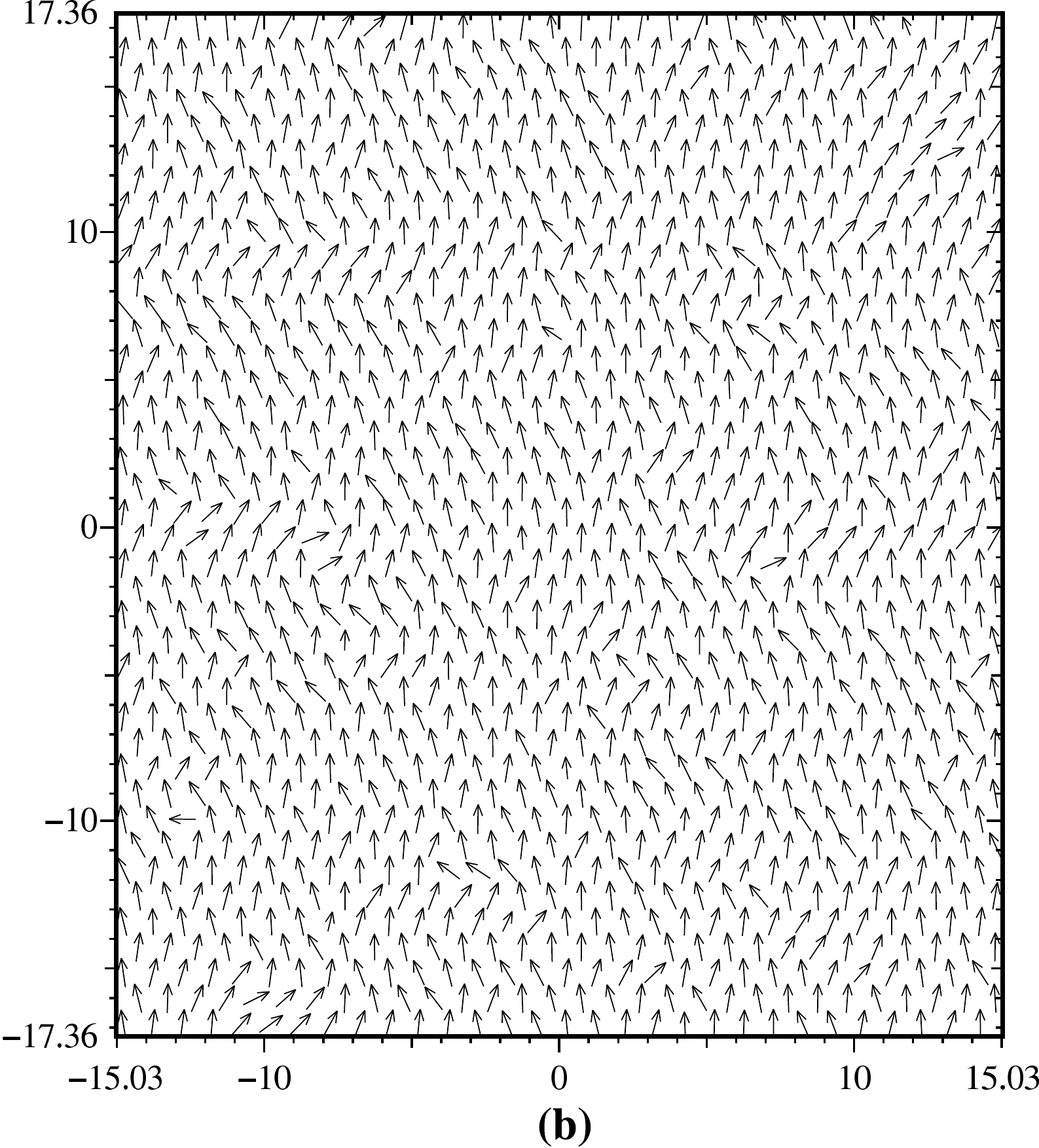}}}
    }
        \caption{Snapshots of  bilayer configurations of particles at close
    packing.  (a) square lattice  ($\rho=1$,  $\mu=2$, $h=1.05$, N=3200);
    (b) hexagonal lattice  ($\rho=1.15$,  $\mu=2$, $h=1.05$, N=2400). 
     The particles in the two layers are on top of each other. The
    arrows denote the projections of the dipole moments on the layer
    plane. (The two layers are shown separately).}
    \label{fig.9}
  \end{center}
\end{figure}

{\centering
\subsection{Structural properties}
}
Structural properties of the bilayer can be conveniently characterized by the coefficients $g^{000}$, $h^{110}$, $h^{112}$ and $h^{220}$ of the expansion of the intra- and interlayer pair correlation functions $h_{intra}(1,2)$ and  $h_{inter}(1,2)$ on a set of rotational invariants as described in subsection 3.3. Selected results for both intra- and interlayer correlation functions for $h=1.05$ at densities $\rho=0.3$ and  $\rho=0.7$ are shown in Figs.\ref{fig.4} - \ref{fig.6}. The intralayer correlation functions for $\mu=1$, reported in  Fig. \ref{fig.4}, agree very well with the correlation functions of the monolayer for the same $\rho$ and for $\mu=1$ (see Fig. 4 of ref.\cite{Lomba:00}).\\
The intralayer correlation functions present a succession of well defined peaks reflecting the formation of chains as also apparent from snapshots of configurations (Figs.\ref{fig.7}(a) and \ref{fig.7}(b)). The peaks sharpen with increasing dipole moment indicating stronger bonding of the particles in the chains. The intralayer correlations appear to be quite insensitive to the layer separation and coincide within statistical error in the range $h=1.05 -2.0$.\\
The interlayer correlation function gives information on the organization of particles in one layer relative to those in the other layer. Although the energy coupling between the layers is quite small one observes a strong correlation of the positional and orientational order of the particles in the two layers (at least for $h < 2$). Inspection of the interlayer distribution function $g^{000}_{inter}$ reveals, for dipole moments $\mu \gtrsim 2$, a high probability of the particles to be on top of each other with opposite directions of the dipole moments ($h^{110}_{inter}$ negative at $s=0$). In addition, at dipole moments $\mu \gtrsim 2.25$, peaks appear in $g^{000}_{inter}$ at  $s=(0.5+n) \sigma$, ($n=0,1,2...$) at which $h^{110}_{inter}$ is positive giving evidence for configurations in which two chains in different layers are nearly on top of each other (possibly some lateral displacement) such that the chain axes of the two chains are displaced by half a HS diameter. In this case dipole moments point in the same direction. The effect is most pronounced at the lower density $\rho=0.3$.\\
The knowledge of $h^{112}_{intra}(s)$ and $h^{112}_{inter}(s)$ enables to recover intralayer and interlayer energies according to
\begin{equation}
\label{v1}
\left\{\begin{array}{ll} 
\displaystyle \frac{\beta {\bar U}^{intra}}{N}&\displaystyle = -\frac{2\pi}{3} \beta\mu^2\rho\int_0^{\infty}\frac{1}{s^2}\rst1 h^{112}_{intra}(s)\rst1ds\\
&\\
\displaystyle \frac{\beta {\bar U}^{inter}}{N}&\displaystyle = -\frac{2\pi}{3} \beta\mu^2\rho\int_0^{\infty}\frac{s}{(s^2+h^2)^{3/2}}\rst1 h^{112}_{inter}(s)\rst1ds
\end{array}\right.
\end{equation}
Similarly, the pressure tensor components are given by 
\begin{equation}
\label{v2}
\left\{\begin{array}{ll} 
\displaystyle {\bar P}_{zz}^{(dd)} &\displaystyle = -4\pi \mu^2\rho^2 h \int_0^{\infty}\frac{s}{(s^2+h^2)^{5/2}}\rst1 h^{112}_{inter}(s)\rst1ds\\
&\\
\displaystyle {\bar \Pi}_{T}^{(dd)} &\displaystyle = -2\pi \mu^2\rho^2\left(\int_0^{\infty}\frac{1}{s^2}\rst1 h^{112}_{intra}(s)\rst1ds+\int_0^{\infty}\frac{s^3}{(s^2+h^2)^{5/2}}\rst1 h^{112}_{inter}(s)\rst1ds\right)
\end{array}\right.
\end{equation}
The quantities ${\bar U}^{intra}$, ${\bar U}^{inter}$, ${\bar P}_{zz}^{(dd)}$ and ${\bar \Pi}_{T}^{(dd)}$  computed with functions $h^{112}_{intra}(s)$ and $h^{112}_{inter}(s)$, can serve as a consistency check with the direct simulation results for energy and pressure using Ewald summations (Tables \ref{tb:Ufixh} and \ref{tb:Pfixh}). Such a comparison is, however, conclusive only if the correlation functions decay to zero on the scale of the simulation box which was only fulfilled at the lower $\mu$ values (cf. Figs.\ref{fig.4} - \ref{fig.6} for the correlation functions). For example at $h=1.05$, $\rho = 0.7$ and  $\mu=1.0$ one has $\beta {\bar U}^{intra}/N=-0.55$, $\beta {\bar U}^{inter}/N=-0.16$, ${\bar P}_{zz}^{(dd)}=-0.43$ and ${\bar \Pi}_{T}^{(dd)}=-1.26$ in good agreement with the results of Tables \ref{tb:Ufixh} and \ref{tb:Pfixh}. For $h=1.05$, $\rho = 0.7$ and $\mu=2.0$, integrating up to half  the box length, one has $\beta {\bar U}^{intra}/N=-5.9$,   $\beta {\bar U}^{inter}/N=-0.42$, ${\bar P}_{zz}^{(dd)}=-1.40$ and ${\bar \Pi}_{T}^{(dd)}=-12.6$ which compares favorably with the values of Tables \ref{tb:Ufixh} and \ref{tb:Pfixh}.\\
Eqs.\ (\ref{v1})-(\ref{v2}), show that  we have the relation $ {\bar \Pi}_{T}^{(dd)}=3\mbox{ }{\bar U}^{intra}/A$ for $h\rightarrow \infty$ ; this asymptotic behavior is in accordance with Eq.(\ref{eos1}). However, it is surprising that Eq.(\ref{eos1}) is verified with such accuracy even for $h=1.05$ (see subsection IV.C and Fig.\ref{fig.3}(a)).\\
The values of $h^{220}$ for $s \gtrsim 7$ agree well with Eq.(\ref{h220limit}). For example, at $\mu=2.5$ on has $P_2 \sim 0.42$ for both densities 0.3 and 0.7. This low value of $P_2$ merely indicates some prevelant local nematic ordering but no global long range  nematic ordering of the dipole moments.\\   
The characterization of the structural organization of the particles in the bilayer at high densities is subject to greater uncertainty due to system size dependence and convergence problems. To illustrate the difficulties we refer to snapshots of configurations at $\rho=0.9$, $\mu=2$ and $h=1.05$ taken at different ``time'' intervals during the MC evolution of the system shown in Figs.\ref{fig.8}(a-d). The system, with $2 \times 1600$ particles, was started from two square lattices with random orientations of the dipole moments. Already after 500 cycles of trial moves small vortices have built up predominantly around particles with dipole moments oriented perpendicularly to the layers (Fig.\ref{fig.8}(a)). As sampling proceeds the vortices grow bigger and large patches develop within which particles arrange with local hexagonal order and parallel alignement of the dipole moments (Fig.\ref{fig.8}(b,c)), clearly an energetically favorable ordering. It remains somewhat unclear whether, for small system sizes, the p.b.c. can stabilize such a ferroelectric arrangement. Such a possibility was indeed observed for a smaller system size ($2 \times 576$ particles) (cf. Fig.\ref{fig.8}(d)), and in one instance ($h=1.005$, $\mu=2$) also for the $ 2 \times 1600$ system though an independent run of similar length ($1 \times 10^6$ cycles) at the same state point retained a vortex arrangement. In some cases, for the smaller $ 2 \times 576 $ system, we also observed formation of stripes with opposite directions of the dipole moments.\\
The structural behavior just described seems typical for dipole strength $\mu \sim 2$ and not to depend much on layer separation in the range $h=1-2$. For larger dipole moments the vortex structure appears to be more stable but, evidently, relaxation of the dipole moments is also slower. For sure is that there  are strong structural correlations between the layers. As for the lower densities, particles arrange preferentially to sit on top of each other with opposite directions of the dipole moments.\\
Finally, in Fig.\ref{fig.9} we show the organization of dipole moments in a bilayer with $h=1.05$ for close packed square and hexagonal lattices of the HS (disks). In both cases the HS in the two layers were taken to be on top of each other. On the square lattices ($\rho=1.0$) the dipole moments in each layer align in parallel lines along the box edges with opposite directions of the dipole moments in neighboring lines (Fig.\ref{fig.9}(a)). A small tendency of microvortex formation is observed. These arrangements are typical of (monolayer) ground state configurations. For a square lattice of in-plane dipoles the ground state is continuously degenerated but thermal contributions can select configurations where rows or colums of parallel spins alternate \cite{Carbog:00}. In contrast, for the 2D triangular lattice with in-plane dipoles, the ground state of the infinite system is ferroelectric \cite{Malozo:91,Rastel:02}; in finite systems the dipolar ordering in the ground state may, however, depend on system size and aspect ratio of the lattice \cite{Politi:06}. In the present finite temperature calculations ($\rho=1.15$) we observe a ferroelectric phase with slight  zigzag ordering of the dipole moments (Fig.\ref{fig.9}(b)). The influence on ordering of dipole strength, system size and use of p.b.c. has still to be investigated. It should be noted also that in our calculations the dipoles are not completely in-plane. As expected, for both lattices, dipole moments in different layers run in opposite directions.

{\centering
\section{SUMMARY AND CONCLUSION}\label{Conclusion}
}
We have investigated by MC simulation the structural and thermodynamic properties of fully orientable dipolar hard spheres mobile in two parallel planar surfaces with particular emphasis on the forces between the two layers. Interlayer correlations turn out to be quite small vanishing practically at layer separations of two HS diameters. The interlayer energy is attractive for all states considered and the normal pressure is negative meaning that an external force must be supplied to keep the layers apart. Indeed isobaric MC simulations, allowing $h$ to fluctuate, did not enable to find an equilibrium state; the system either collapsed (at low applied negative pressure) or the two layers drifted away (at larger pressures). The normal pressure is well described by a $-1/h^{5}$ dependence at larger separations in agreement with a second order perturbation theory of the interaction free energy of the surfaces in an infinite dielectric medium by Attard and Mitchell \cite{Attard:87,Attard:88}. Despite the weak interlayer energy there are strong correlations for the structural behavior of the particles in the two layers. Particles preferentially sit on top of each other with opposite orientations of the dipole moments. At densities of the order $\rho \sim 0.9$ convergence of the MC
sampling is slow and, moreover, finite size effects may affect the results. Although we believe that for large systems vortex formation is the preferred structure, arrangements with ferroelectric ordering or stripes with up and down orientations of the dipole moments were stabilized in the smaller systems, likely by the use of periodic boundary conditions. These problems clearly need a more detailed investigation.\\
As an extension of the present work it would be of interest to consider the case where the media on either side of the layers have different dielectric constants, as would be the case, for instance, in a lipid bilayer model where the hydrocarbon tails and aquous regions are approximated by ideal dielectrics. Although the surface polarization arising from the dielectric discontinuities can in principle be taken into account through dielectric images \cite{Jonsso:83} few simulation results have been presented so far \cite{Granfe:88}. Such simulations could valuably add to the comprehension of the origin of the repulsive "hydration" forces measured in phospholipid bilayers at short distances \cite{Rand:89}. Existing theoretical approaches based on continuum electrostatics \cite{Attard:88,Jonsso:88} seem to fail to predict correctly these repulsive forces.

\begin{center}
\large{\bf ACKNOWLEDGEMENTS}
\end{center}
The computations have been performed on IBM Regatta Power 4 stations of IDRIS ({\it Institut du D\'eveloppement et des Ressources en Informatique Scientifique\/}) under projects 0672104 and 0682104. C. Alvarez acknowledges financial support by COLCIENCIAS and SECAB (Executive Secretariat of the Andes Bello Convention) in the framework of the cooperation treaty 065-2002. The work has benefitted from support of project ECOS-Nord CO5PO2.

\newpage
\begin{center}
\large{\bf APPENDIX A: EWALD SUMS FOR THE DIPOLAR ENERGY OF THE BILAYER}
\end{center}
\renewcommand{\theequation}{A.\arabic{equation}}
\setcounter{equation}{0}
The total dipolar energy of the bilayer computed with the Ewald method is written  as
\begin{equation}
\displaystyle U_{dd} = E_{\bm{r}}+E^{(1)}_{\bm{G}\neq
  0}+E^{(2)}_{\bm{G}\neq 0}+E^{(3)}_{\bm{G}\neq 0}+E_{\bm{G}= 0}.
\label{B1}
\end{equation}
Here $E_{\bm{r}}$ is the short range (direct space) contribution to the energy given by
\begin{equation}
\displaystyle E_{\bm{r}}=\frac{1}{2}\sum_{i\neq j}
\mbox{\large{[}}(\bm{\mu}_i\cdot\bm{\mu}_j)\mbox{B}(r_{ij})-(\bm{\mu}_i\cdot\bm{r}_{ij})(\bm{\mu}_j\cdot\bm{r}_{ij})\mbox{C}(r_{ij})
  \mbox{\large{]}}
\label{B2}
\end{equation}
with
\begin{equation}
\label{B3}
\left\{\begin{array}{ll} 
\displaystyle \mbox{B}(r) &\displaystyle = \frac{\mbox{erfc}(\alpha r)}{r^3}+\frac{2\alpha}{\sqrt{\pi}}\frac{\exp(-\alpha^2r^2)}{r^2}\\
&\\
\displaystyle \mbox{C}(r) &\displaystyle = 3\frac{\mbox{erfc}(\alpha r)}{r^5}+\frac{2\alpha}{\sqrt{\pi}}\mbox{\large{(}}2\alpha^2+\frac{3}{r^2}\mbox{\large{)}}\frac{\exp(-\alpha^2r^2)}{r^2}
\end{array}\right.
\end{equation}
In Eq.(\ref{B2}) it is assumed that the parameter $\alpha$ is sufficiently large to restrict interactions to the basic simulation cell.  The energy $E_{\bm{r}}$ can, in turn, be  separated into an intralayer $E_{\bm r}^{intra}$ and an interlayer $E_{\bm r}^{inter}$ contribution. The four last terms in Eq.(\ref{B1}) are the reciprocal space contributions. Each of the terms is again separated into intralayer and interlayer contributions. They are split into three contributions: $E^{(1)}_{\bm{G}\neq 0}$ involves only  coupling between the normal components of dipole moments, $E^{(2)}_{\bm{G}\neq 0}$ coupling between in-plane and normal components of dipoles and $E^{(3)}_{\bm{G}\neq 0}$  in-plane coupling. 
Contributions to the interlayer energy are given by
\begin{equation}
\label{B4}
\left\{\begin{array}{ll} 
\displaystyle E^{(1, inter)}_{\bm{G}\neq 0}&\displaystyle= \frac{\pi}{A}\sum_{\bm{G}\neq 0}\mbox{I}(\alpha,G;h)\times  \Re e\mbox{\Large{[}} \mbox{\large{(}}\sum_{i\in L_1} \mu_i^{z}\exp(i\bm{G}\cdot\bm{s}_i)\mbox{\large{)}} \mbox{\large{(}}\sum_{j\in L_2} \mu_j^{z}\exp(-i\bm{G}\cdot\bm{s}_j)\mbox{\large{)}}\mbox{\Large{]}}  \\
&\\
\displaystyle E^{(2,inter)}_{\bm{G}\neq 0}&\displaystyle= \frac{\pi}{A}\sum_{\bm{G}\neq 0}\mbox{J}(\alpha,G;h)\times \Im m\mbox{\Large{[}}\mbox{\large{(}}\sum_{i\in L_1} (\bm{\mu}_i\cdot\bm{G})\exp(i\bm{G}\cdot\bm{s}_i)\mbox{\large{)}}\mbox{\large{(}}\sum_{j\in L_2} \mu_j^{z}\exp(-i\bm{G}\cdot\bm{s}_j)\mbox{\large{)}} \\
&\\
&\displaystyle +\mbox{\large{(}}\sum_{i\in L_1} \mu_i^{z}\exp(i\bm{G}\cdot\bm{s}_i)\mbox{\large{)}} \mbox{\large{(}}\sum_{j\in L_2} (\bm{\mu}_j\cdot\bm{G})\exp(-i\bm{G}\cdot\bm{s}_j)\mbox{\large{)}}\mbox{\Large{]}} \\
&\\
\displaystyle E^{(3, inter)}_{\bm{G}\neq 0}&\displaystyle=\frac{\pi}{A}\sum_{\bm{G}\neq 0}\mbox{K}(\alpha,G;h) \times  \Re e\mbox{\Large{[}} \mbox{\large{(}}\sum_{i\in L_1} (\bm{\mu}_i\cdot\bm{G})\exp(i\bm{G}\cdot\bm{s}_i)\mbox{\large{)}} \mbox{\large{(}}\sum_{j\in L_2} (\bm{\mu}_j\cdot\bm{G})\exp(-i\bm{G}\cdot\bm{s}_j)\mbox{\large{)}}\mbox{\Large{]}}
\end{array}\right.
\end{equation}
where $\Re e[z]$ and $\Im m[z]$ are the real and imaginary parts of the complex number $z$, respectively. $\bm{G}= 2 \pi \big( \frac{n_x}{L_x},\frac{n_y}{L_y} \big)$, ($n_x$,$n_y$ integers) is a two-dimensional vector in recriprocal lattice and $G=\parallel \bm{G} \parallel$. The functions $\mbox{I}(\alpha,G;h)$, $\mbox{J}(\alpha,G;h)$ and $\mbox{K}(\alpha,G;h)$ are given by
\begin{equation}
\label{B5}
\left\{\begin{array}{ll} 
\displaystyle \mbox{I}(\alpha,G;h)&\displaystyle= \frac{4\alpha}{\sqrt{\pi}}\exp\mbox{\large{(}}-\frac{G^2}{4\alpha^2}-\alpha^2h^2\mbox{\large{)}}-G^2\mbox{K}(\alpha,G;h)\\
&\\
\displaystyle \mbox{J}(\alpha,G;h)&\displaystyle=\exp(Gh)\mbox{erfc}\mbox{\large{(}}\frac{G}{2\alpha}+\alpha h\mbox{\large{)}}-\exp(-Gh)\mbox{erfc}\mbox{\large{(}}\frac{G}{2\alpha}-\alpha h\mbox{\large{)}}\\
&\\
\displaystyle \mbox{K}(\alpha,G;h)&\displaystyle=\frac{1}{G}\mbox{\Large{[}} \exp(Gh)\mbox{erfc}\mbox{\large{(}}\frac{G}{2\alpha}+\alpha h\mbox{\large{)}}+\exp(-Gh)\mbox{erfc}\mbox{\large{(}}\frac{G}{2\alpha}-\alpha h\mbox{\large{)}}\mbox{\Large{]}}
\end{array}\right.
\end{equation}
The constant term is
\begin{equation}
\label{B6}
\displaystyle E_{\bm{G}= 0}^{(inter)}=\frac{4\alpha\sqrt{\pi}}{A}\exp(-\alpha^2h^2)\mbox{\Large{[}}\mbox{\large{(}}\sum_{i\in L_1} \mu_i^{z}\mbox{\large{)}}\mbox{\large{(}}\sum_{j\in L_2} \mu_j^{z}\mbox{\large{)}}\mbox{\Large{]}}
\end{equation}
Contributions to intralayer the energy are given by
\begin{equation}
\label{B7}
\left\{\begin{array}{ll} 
\displaystyle E^{(1, intra)}_{\bm{G}\neq 0}&\displaystyle=\frac{\pi}{A}\sum_{\bm{G}\neq 0}\mbox{D}(\alpha,G)\mbox{\Large{[}} \mid \sum_{i\in L_1} \mu_i^{z}\exp(i\bm{G}\cdot\bm{s}_i)\mid^2+\mid \sum_{j\in L_2} \mu_j^{z}\exp(i\bm{G}\cdot\bm{s}_j)\mid^2\mbox{\Large{]}}\\
&\\
\displaystyle E^{(2, intra)}_{\bm{G}\neq 0}&\displaystyle=0\\
&\\
\displaystyle E^{(3, intra)}_{\bm{G}\neq 0}&\displaystyle=\frac{\pi}{A}\sum_{\bm{G}\neq 0}\mbox{H}(\alpha,G)\mbox{\Large{[}} \mid \sum_{i\in L_1} (\bm{\mu}_i\cdot\bm{G})\exp(i\bm{G}\cdot\bm{s}_i)\mid^2+\mid \sum_{j\in L_2} (\bm{\mu}_j\cdot\bm{G})\exp(i\bm{G}\cdot\bm{s}_j)\mid^2\mbox{\Large{]}}
\end{array}\right.
\end{equation}
with
\begin{equation}
\label{B8}
\left\{\begin{array}{ll} 
\displaystyle \mbox{D}(\alpha,G)&\displaystyle=\frac{2\alpha}{\sqrt{\pi}} \exp(-G^2/4\alpha^2) -G\rst1\mbox{erfc}(G/2\alpha)\\
&\\
\displaystyle\mbox{H}(\alpha,G)&\displaystyle=\frac{\mbox{erfc}(G/2\alpha)}{G}
\end{array}\right.
\end{equation}
and the constant is
\begin{equation}
\label{B9}
\displaystyle E_{\bm{G}= 0}^{(\mbox{\footnotesize intra})}=\frac{2\alpha\sqrt{\pi}}{A}\mbox{\Large{[}}\mbox{\large{(}}\sum_{i\in L_1} \mu_i^{z}\mbox{\large{)}}^2+\mbox{\large{(}}\sum_{j\in L_2} \mu_j^{z}\mbox{\large{)}}^2\mbox{\Large{]}}-\frac{2\alpha^3}{3\sqrt{\pi}}\sum_i \bm{\mu}_i^2.
\end{equation}
Due to the 2d character of $\bm{G}$ it is easily seen from the corresponding term in Eq.(\ref{B4}) (interlayer contribution) that  $E^{(2, intra)}_{\bm{G}\neq 0}$ must vanish.

\newpage
\begin{center}
\large{\bf APPENDIX B: THE MICROSCOPIC STRESS TENSOR OF THE BILAYER}
\end{center}
\renewcommand{\theequation}{B.\arabic{equation}}
\setcounter{equation}{0}
In this Appendix, we derive the microscopic stress tensor for the bilayer system from its equations of motion, in a  way similar  to the one of ref.\cite{Schofield:82}(a) for inhomogeneous fluids. The microscopic stress tensor of the bilayer is split into normal $\sigma_N$ and lateral $\sigma_T$ components as
\begin{equation}
\label{BB1}
\displaystyle \sigma=\sigma_T+\sigma_N=\left(\begin{array}{lll} \sigma_{xx} & \sigma_{xy} & 0\\ \sigma_{xy} & \sigma_{yy} & 0 \\ 0 & 0 & 0\end{array}\right)+\left(\begin{array}{lll} 0 & 0 &  \sigma_{xz}  \\ 0 & 0 & \sigma_{yz} \\ \sigma_{xz} & \sigma_{yz} & \sigma_{zz} \end{array}\right)
\end{equation}
The Lagrangian function of the bilayer system, with the constraints $z_i=H_1$, for $i\in L_1$, and $z_i=H_2$, for $i\in L_2$ is given by
\begin{equation}
\label{BB2}
\begin{array}{ll} 
\displaystyle\mathcal{L} &\displaystyle =\sum_{i\in L_1\cup L_2}\frac{1}{2}m_i\dot{\bm{s}_i}^2+\sum_{i\in L_1}\frac{1}{2}m_i\dot{H_1}^2+\sum_{i\in L_2}\frac{1}{2}m_i\dot{H_2}^2\\
&\\
&\displaystyle -\frac{1}{2}\sum_i\sum_{j\neq i}\Phi(\bm{s}_{ij},z_{ij})-\sum_{i\in L_1\cup L_2}\Phi_{ext}(\bm{s}_{i},z_{i})
\end{array}
\end{equation}
where $\Phi$ is the pair potential energy due to interactions between particles and $\Phi_{ext}$ represents the action of any external fields. In the above equation, $H_1$ and $H_2$ are collective variables associated with the z-coordinate of the layers. From the lagrangian of the system, we obtain the equations of motion for the particles in the layer $L_1$ and the collective variable $H_1$ : 
\begin{equation}
\label{BB3}
\begin{array}{ll} 
m\ddot{\bm{s}_i}&\displaystyle = -\sum_{j\in L_1, j\neq i}\bm{\nabla}_{i}\Phi(\bm{s}_{ij},0)-\sum_{j\in L_2}\bm{\nabla}_{i}\Phi(\bm{s}_{ij},H_2-H_1)-\bm{\nabla}_{i}\Phi_{ext}(\bm{s}_{i},H_1)
\end{array}
\end{equation}
\begin{equation}
\label{BB4}
\begin{array}{ll} 
N_0m\ddot{H_1}&\displaystyle = \frac{\partial}{\partial z}\sum_{i\in L_1}\sum_{j\in L_2}\Phi(\bm{s}_{ij},H_2-H_1)-\frac{\partial}{\partial z}\sum_{i\in L_1}\Phi_{ext}(\bm{s}_{i},H_1)
\end{array}
\end{equation}
and similar equations for the layer $L_2$. $m$ denotes the mass of the particles.\\
The momentum density for the bilayer system can be written as
\begin{equation}
\label{BB5}
\begin{array}{ll} 
\displaystyle \bm{J}(\bm{s},z,t)&\displaystyle =\bm{J}_T(\bm{s},z,t)+J_N(\bm{s},z,t)\hat{\bm{e}}_z\\
&\\
&\displaystyle = m\mbox{ }\delta(z-H_1)\sum_{i\in L_1}\dot{\bm{s}_i}\mbox{ }\delta(\bm{s}-\bm{s}_i)+m\mbox{ }\delta(z-H_2)\sum_{i\in L_2}\dot{\bm{s}_i}\mbox{ }\delta(\bm{s}-\bm{s}_i)\\
&\\
&\displaystyle + m\dot{H_1}\mbox{ }\delta(z-H_1)\sum_{i\in L_1}\delta(\bm{s}-\bm{s}_i)\hat{\bm{e}}_z+ m\dot{H_2}\mbox{ }\delta(z-H_2)\sum_{i\in L_2}\delta(\bm{s}-\bm{s}_i)\hat{\bm{e}}_z
\end{array}
\end{equation}
where $\delta(x)$ is the Dirac distribution. From the time derivative of the momentum density, we obtain easily \cite{Schofield:82} the kinetic contribution to the lateral component of the stress tensor as
\begin{equation}
\label{BB6}
\displaystyle \sigma_{\alpha\beta}^K(\bm{s},z,t)=-m\mbox{ }\delta(z-H_1)\sum_{i\in L_1}\dot{s_i}^{\alpha}\dot{s_i}^{\beta}\mbox{ }\delta(\bm{s}-\bm{s}_i)-m\mbox{ }\delta(z-H_2)\sum_{i\in L_2}\dot{s_i}^{\alpha}\dot{s_i}^{\beta}\mbox{ }\delta(\bm{s}-\bm{s}_i)
\end{equation}
with $\alpha,\beta=x,y$. 
The kinetic contribution to the normal component is obtained similarly as
\begin{equation}
\label{BB7}
\left\{ \begin{array}{ll} 
\displaystyle \sigma_{\alpha z}^K(\bm{s},z,t) &\displaystyle =-m\dot{H_1}\mbox{ }\delta(z-H_1)\sum_{i\in L_1}\dot{s_i}^{\alpha}\mbox{ }\delta(\bm{s}-\bm{s}_i)-m\dot{H_2}\mbox{ }\delta(z-H_2)\sum_{i\in L_2}\dot{s_i}^{\alpha}\mbox{ }\delta(\bm{s}-\bm{s}_i)\\
&\\
\displaystyle \sigma_{z z}^K(\bm{s},z,t) &\displaystyle = - m\dot{H_1}^2\mbox{ }\delta(z-H_1)\sum_{i\in L_1}\mbox{ }\delta(\bm{s}-\bm{s}_i)- m\dot{H_2}^2\mbox{ }\delta(z-H_2)\sum_{i\in L_2}\mbox{ }\delta(\bm{s}-\bm{s}_i)
\end{array}\right.
\end{equation}
The configurational contributions to the stress tensor, follow from Eq.(\ref{BB3})
\begin{equation}
\label{BB8}
\begin{array}{ll} 
\displaystyle \sigma_{\alpha\beta}^C(\bm{s},z,t)&\displaystyle = \left[ \frac{1}{2}\sum_{i\in L_1}\sum_{j\in L_1, j\neq i}\nabla_{i}^{\alpha}\Phi(\bm{s}_{ij},0)\int_{C_{ij}}dl^{\beta}\mbox{ } \delta(\bm{s}-\bm{l})\right. \\
&\\
&\displaystyle\left . +\frac{1}{2}\sum_{i\in L_1}\sum_{j\in L_2}\nabla_{i}^{\alpha}\Phi(\bm{s}_{ij},H_2-H_1)\int_{C_{ij}}dl^{\beta}\mbox{ } \delta(\bm{s}-\bm{l})\right] \mbox{ }\delta(z-H_1)\\
&\\
&\displaystyle +\left[ \frac{1}{2}\sum_{i\in L_2}\sum_{j\in L_2, j\neq i}\nabla_{i}^{\alpha}\Phi(\bm{s}_{ij},0)\int_{C_{ij}}dl^{\beta}\mbox{ } \delta(\bm{s}-\bm{l})\right.\\
&\\
&\displaystyle \left.+\frac{1}{2}\sum_{i\in L_2}\sum_{j\in L_1}\nabla_{i}^{\alpha}\Phi(\bm{s}_{ij},H_1-H_2)\int_{C_{ij}}dl^{\beta}\mbox{ } \delta(\bm{s}-\bm{l})\right]  \mbox{ }\delta(z-H_2)
\end{array}
\end{equation}
with $\alpha=x,y$ and $C_{ij}$ a contour joining $\bm{s}_i$ to $\bm{s}_j$ in the plane perpendicular to the z direction. Eqs.(\ref{BB8}) and (\ref{BB6}) allow to fully determine the lateral component of the stress tensor of the bilayer. The integrals in Eq.(\ref{BB8}) can be evaluated by using the parametrization proposed by Irving and Kirkwood \cite{Schofield:82}(b), namely
\begin{equation}
\label{BB9}
\displaystyle\sum_{i\in L_1}\sum_{j\in L_1, j\neq i}\nabla_{i}^{\alpha}\Phi(\bm{s}_{ij},0)\int_{C_{ij}}dl^{\beta}\mbox{ } \delta(\bm{s}-\bm{l})\displaystyle = \sum_{i\in L_1}\sum_{j\in L_1, j\neq i}s_{ij}^{\beta}\nabla_{i}^{\alpha}\Phi(\bm{s}_{ij},0)\int_0^1d\lambda\mbox{ }\delta(\bm{s}-\lambda\bm{s}_j-(1-\lambda)\bm{s}_i)
\end{equation}
and
\begin{equation}
\label{BB10}
\begin{array}{lll} 
&\displaystyle \sum_{i\in L_1}\sum_{j\in L_2}\nabla_{i}^{\alpha}\Phi(\bm{s}_{ij},H_2-H_1)\int_{C_{ij}}dl^{\beta}\mbox{ } \delta(\bm{s}-\bm{l})&\\
&&\\
& \displaystyle =\sum_{i\in L_1}\sum_{j\in L_2} s_{ij}^{\beta}\nabla_{i}^{\alpha}\Phi(\bm{s}_{ij},H_2-H_1)\int_0^1d\lambda\mbox{ }\delta(\bm{s}-\lambda\bm{s}_j-(1-\lambda)\bm{s}_i)&\\
&&\\
&\displaystyle =\sum_{i\in L_2}\sum_{j\in L_1} s_{ij}^{\beta}\nabla_{i}^{\alpha}\Phi(\bm{s}_{ij},H_1-H_2)\int_0^1d\lambda\mbox{ }\delta(\bm{s}-\lambda\bm{s}_j-(1-\lambda)\bm{s}_i)&
\end{array}
\end{equation}
Eqs.(\ref{BB6}) and (\ref{BB8}) show that $\sigma_{\alpha \beta}$ can be written in
the form  ($\alpha,\beta = x,y$)
\begin{equation}
\label{BB11}
\sigma_{\alpha \beta}(\bm{s},z,t)=\tau_{\alpha \beta}^{(1)}(\bm{s},t)\mbox{ }\delta(z-H_1)+\tau_{\alpha \beta}^{(2)}(\bm{s},t)\mbox{ }\delta(z-H_2)
\end{equation}
One should note that, if $z\neq H_1$ and $z\neq H_2$ then $\sigma_{\alpha \beta}(\bm{s},z,t)=0$.\\  
In accord with solid surface physics  we define the surface stress tensor as
\begin{equation}
\label{BB12}
\eta_{\alpha \beta}(\bm{s},t)=\int\sigma_{\alpha \beta}(\bm{s},z,t)dz= \tau_{\alpha \beta}^{(1)}(\bm{s},t)+\tau_{\alpha \beta}^{(2)}(\bm{s},t)
\end{equation}
If one adopts the two-component monolayer picture discussed in the main text, then each contribution $\tau_{\alpha \beta}^{(1)}$ and $\tau_{\alpha \beta}^{(2)}$ correspond respectively to partial contribution of each species to the surface stress tensor.\\
From the surface stress tensor we define the lateral component of the pressure tensor of the bilayer as the ensemble average of the surface stress tensor  as
\begin{equation}
\label{BB13}
\Pi_{\alpha\beta}=-\Big<\frac{1}{A}\int_{L_1\cup L_2}d\bm{s}\mbox{ }\eta_{\alpha \beta}(\bm{s},t)\Big>.
\end{equation}
It follows that
\begin{equation}
\label{BB14}
\begin{array}{ll} 
\displaystyle \Pi_{\alpha\beta}= 2\rho kT\delta_{\alpha\beta} &\displaystyle- \Big<\frac{1}{2A}\sum_{i\in L_1}\sum_{j\in L_1, j\neq i}s_{ij}^{\beta}\nabla_{i}^{\alpha}\Phi(\bm{s}_{ij},0)\Big>\\
&\\
&\displaystyle- \Big<\frac{1}{2A}\sum_{i\in L_2}\sum_{j\in L_2, j\neq i}s_{ij}^{\beta}\nabla_{i}^{\alpha}\Phi(\bm{s}_{ij},0)\Big>\\
&\\
&\displaystyle -\Big<\frac{1}{A}\sum_{i\in L_1}\sum_{j\in L_2} s_{ij}^{\beta}\nabla_{i}^{\alpha}\Phi(\bm{s}_{ij},H_2-H_1)\Big>.
\end{array}
\end{equation}
The average lateral pressure $\Pi_T$ and the surface stress $\tilde{\eta}$ are then given by
\begin{equation}
\label{BB15}
\Pi_T=\frac{1}{2}(\Pi_{xx}+\Pi_{yy})=-\tilde{\eta}
\end{equation}
The configurational contribution to the normal component $\sigma_{zz}$ allows to obtain the force acting on the layers. From the equations of motion of $H_1$ and $H_2$, we obtain
\begin{equation}
\label{BB16}
\begin{array}{ll} 
\displaystyle \frac{\partial}{\partial z} \sigma_{zz}^C(\bm{s},z,t) &\displaystyle= \frac{1}{N_0}\mbox{\Large{(}}\sum_{n\in L_1}\mbox{ } \delta(\bm{s}-\bm{s}_n)\mbox{\Large{)}}\mbox{ }\delta(z-H_1)\mbox{\Large{(}}\frac{\partial}{\partial z}\sum_{i\in L_1}\sum_{j\in L_2}\left.\Phi(\bm{s}_{ij},z)\right|_{z=H_2-H_1}\mbox{\Large{)}}\\
&\\
&\displaystyle -\frac{1}{N_0}\mbox{\Large{(}}\sum_{n\in L_2}\mbox{ } \delta(\bm{s}-\bm{s}_n)\mbox{\Large{)}}\mbox{ }\delta(z-H_2)\mbox{\Large{(}}\frac{\partial}{\partial z}\sum_{i\in L_1}\sum_{j\in L_2}\left.\Phi(\bm{s}_{ij},z)\right|_{z=H_2-H_1}\mbox{\Large{)}}.
\end{array}
\end{equation}
Thus, the total force $F_{2\rightarrow 1}^{z}$ acting on layer $L_1$ due to the particles in layer $L_2$ is given  by
\begin{equation}
\label{BB17}
\displaystyle F_{2\rightarrow 1}^{z}=-\int d\bm{s} \frac{\partial}{\partial z} \sigma_{zz}^C(\bm{s},z=H_1,t)=-\frac{\partial}{\partial z}\sum_{i\in L_1}\sum_{j\in L_2}\left.\Phi(\bm{s}_{ij},z)\right|_{z=H_2-H_1}
\end{equation}
and, obviously, we have 
\begin{equation}
\label{BB18}
\displaystyle F_{1\rightarrow 2}^{z}=-\int d\bm{s} \frac{\partial}{\partial z} \sigma_{zz}^C(\bm{s},z=H_2,t)=-F_{2\rightarrow 1}^{z}
\end{equation}
The average force by unit area is 
\begin{equation}
\label{BB19}
\displaystyle f_{2\rightarrow 1}^{z}=\Big<\frac{1}{A}F_{2\rightarrow 1}^{z}\Big> =-\Big<\frac{1}{A}\frac{\partial}{\partial z}\sum_{i\in L_1}\sum_{j\in L_2}\left.\Phi(\bm{s}_{ij},z)\right|_{z=H_2-H_1}\Big>=P_{zz}=P_N
\end{equation}
The equation (\ref{BB19}) for $P_N$ is in full agreement with the derivation of the normal pressure derived for similar systems in refs.\cite{Schofield:82,Walton:85,Klapp:02,Widom:82}.\\
If the z-coordinates of the layers are fixed, as is the case in most of the computations in the present work, an external field compensates exactly the microscopic forces. In this case we have $H_1=-H_2=h/2$, $\dot{H}_1= \dot{H}_2=0$ and $\ddot{H}_1= \ddot{H}_2=0$ and the external forces are given by
\begin{equation}
\label{BB20}
\displaystyle F_{ext,1}^z=\sum_{i\in L_1}\frac{\partial}{\partial z}\Phi_{ext}(\bm{s}_{i},\frac{h}{2})=-F_{2\rightarrow 1}^{z}
\end{equation}
and
\begin{equation}
\label{BB21}
\displaystyle F_{ext,2}^z=-F_{1\rightarrow 2}^{z}=F_{2\rightarrow 1}^{z}=-F_{ext,1}^z
\end{equation}

\newpage
\begin{center}
\large{\bf APPENDIX C: RECIPROCAL SPACE CONTRIBUTIONS TO THE PRESSURE TENSOR AND FORCES}
\end{center}
\renewcommand{\theequation}{C.\arabic{equation}}
\setcounter{equation}{0}

The general formulae for the components of the stress tensor in terms of the interaction potential are given in section 2. In this appendix, we give explicit expressions for the reciprocal space contribution in an Ewald sum  of the stress tensor components. They can be obtained directly from the results of Appendix A or from the general derivation given by Heyes \cite{Heyes:94} for quasi-two dimensional systems.\\
The short ranged contributions are easily obtained from Eqs.(A.2-3).\\
From Eq.(\ref{press1}) and with notations of Appendix A, we have, for the bilayer system,
\begin{equation}
\label{C1}
\Pi_{T}^{(dd,G)}=-\frac{1}{2A}\Big<\sum_i\bm{s}_i\cdot\nabla_{\bm{s}_i}(E_{{\bm G}\neq 0}^{(intra)}+E_{{\bm G}\neq 0}^{(inter)})\Big>
\end{equation}
\begin{equation}
\label{C2}
P_{zz}^{(dd,G)}=-\frac{1}{A}\Big< \left.\frac{\partial}{\partial z}E_{{\bm G}\neq 0}^{(inter)}\right|_{z=h} \Big>
\end{equation}
The intralayer contributions to the lateral components of the stress tensor are given by
\begin{equation}
\label{C3}
\begin{array}{ll}
\displaystyle \sum_i\bm{s}_i\cdot \nabla_{\bm{s}_i}E_{{\bm G}\neq 0}^{(1,intra)}&\displaystyle= -\frac{2\pi}{A}\sum_{\bm G\neq 0}\mbox{D}(\alpha,G)\\
&\\
&\displaystyle \times\Im m\mbox{\Large{[}}\mbox{\large{(}}\sum_{i\in L_1} (\bm{G}\cdot\bm{s}_i)\mu_i^{z}\exp(i\bm{G}\cdot\bm{s}_i)\mbox{\large{)}}\mbox{\large{(}}\sum_{i\in L_1}\mu_i^{z}\exp(i\bm{G}\cdot\bm{s}_i)\mbox{\large{)}}\\
&\\
&\displaystyle+\mbox{\large{(}}\sum_{i\in L_2} (\bm{G}\cdot\bm{s}_i)\mu_i^{z}\exp(i\bm{G}\cdot\bm{s}_i)\mbox{\large{)}}\mbox{\large{(}}\sum_{i\in L_2}\mu_i^{z}\exp(i\bm{G}\cdot\bm{s}_i)\mbox{\large{)}}\mbox{\Large{]}}\\
\end{array}
\end{equation}
\begin{quote}
$\displaystyle \sum_i\bm{s}_i\cdot \nabla_{\bm{s}_i}E_{{\bm G}\neq 0}^{(2,intra)}\displaystyle= 0$ \hspace{4.in} (C.4)\\
\end{quote}
\setcounter{equation}{4}
\begin{equation}
\label{C5}
\begin{array}{ll}
\displaystyle \sum_i\bm{s}_i\cdot \nabla_{\bm{s}_i}E_{{\bm G}\neq 0}^{(3,intra)}&\displaystyle= -\frac{2\pi}{A}\sum_{\bm G\neq 0}\mbox{H}(\alpha,G)\\
&\\
&\displaystyle \times\Im m\mbox{\Large{[}}\mbox{\large{(}}\sum_{i\in L_1} (\bm{G}\cdot\bm{s}_i)(\bm{\mu}_i\cdot\bm{G})\exp(i\bm{G}\cdot\bm{s}_i)\mbox{\large{)}}\mbox{\large{(}}\sum_{i\in L_1}(\bm{\mu}_i\cdot\bm{G})\exp(i\bm{G}\cdot\bm{s}_i)\mbox{\large{)}}\\
&\\
&\displaystyle+\mbox{\large{(}}\sum_{i\in L_2} (\bm{G}\cdot\bm{s}_i)(\bm{\mu}_i\cdot\bm{G})\exp(i\bm{G}\cdot\bm{s}_i)\mbox{\large{)}}\mbox{\large{(}}\sum_{i\in L_2}(\bm{\mu}_i\cdot\bm{G})\exp(i\bm{G}\cdot\bm{s}_i)\mbox{\large{)}}\mbox{\Large{]}}
\end{array}
\end{equation}
with functions $D$ and $H$ as defined in Eq.(\ref{B8}).\\
Interlayer contributions are given by
\begin{equation}
\label{C6}
\begin{array}{ll}
\displaystyle \sum_i\bm{s}_i\cdot \nabla_{\bm{s}_i}E_{{\bm G}\neq 0}^{(1,inter)}&\displaystyle =\frac{\pi}{A}\sum_{{\bm G}\neq 0}\mbox{I}(\alpha,G;h)\\
&\\
&\displaystyle \times \Im m\mbox{\Large{[}}\mbox{\large{(}}\sum_{i\in L_1}\mu_i^z\mbox{exp}(i{\bm G}\cdot{\bm s}_i)\mbox{\large{)}}\mbox{\large{(}}\sum_{j\in L_2}\mu_j^z({\bm G}\cdot {\bm s}_j)\mbox{exp}(-i{\bm G}\cdot{\bm s}_j)\mbox{\large{)}}\\
&\\
&\displaystyle -\mbox{\large{(}}\sum_{i\in L_1}\mu_i^z({\bm G}\cdot {\bm s}_i)\mbox{exp}(i{\bm G}\cdot {\bm s}_i)\mbox{\large{)}}\mbox{\large{(}}\sum_{j\in L_2}\mu_j^z\mbox{exp}(-i{\bm G}\cdot {\bm s}_j)\mbox{\large{)}}\mbox{\Large{]}}
\end{array}
\end{equation}
\begin{equation}
\label{C7}
\begin{array}{ll}
\displaystyle \sum_i\bm{s}_i\cdot \nabla_{\bm{s}_i}E_{{\bm G}\neq 0}^{(2,inter)}&\displaystyle = \frac{\pi}{A}\sum_{{\bm G}\neq 0}\mbox{J}(\alpha,G;h)\\
&\\
&\displaystyle \times \Re e \mbox{\Large{[}}\mbox{\large{(}}\sum_{i\in L_1}\mu_i^z({\bm G}\cdot{\bm s}_i)\mbox{exp}(i{\bm G}\cdot {\bm s}_i)\mbox{\large{)}}\mbox{\large{(}}\sum_{j\in L_2}({\bm\mu}_j\cdot {\bm G})\mbox{exp}(-i{\bm G}\cdot {\bm s}_j)\mbox{\large{)}}\\
&\\
&\displaystyle - \mbox{\large{(}}\sum_{i\in L_1}\mu_i^z\mbox{exp}(i{\bm G}\cdot {\bm s}_i)\mbox{\large{)}}\mbox{\large{(}}\sum_{j\in L_2}({\bm\mu}_j\cdot {\bm G})({\bm G}\cdot{\bm s}_j)\mbox{exp}(-i{\bm G}\cdot {\bm s}_j)\mbox{\large{)}}\\
&\\
&\displaystyle + \mbox{\large{(}}\sum_{i\in L_1}({\bm\mu}_i\cdot {\bm G})({\bm G}\cdot{\bm s}_i)\mbox{exp}(i{\bm G}\cdot{\bm s}_i)\mbox{\large{)}}\mbox{\large{(}}\sum_{j\in L_2}\mu_j^z\mbox{exp}(-i{\bm G}\cdot {\bm s}_j)\mbox{\large{)}}\\
&\\
&\displaystyle - \mbox{\large{(}}\sum_{i\in L_1}({\bm\mu}_i\cdot {\bm G})\mbox{exp}(i{\bm G}\cdot{\bm s}_i)\mbox{\large{)}}\mbox{\large{(}}\sum_{j\in L_2}\mu_j^z({\bm G}\cdot{\bm s}_j)\mbox{exp}(-i{\bm G}\cdot{\bm s}_j)\mbox{\large{)}}\mbox{\Large{]}}
\end{array}
\end{equation}
\begin{equation}
\label{C8}
\begin{array}{ll}
\displaystyle \sum_i\bm{s}_i\cdot \nabla_{\bm{s}_i} E_{{\bm G}\neq 0}^{(3,inter)}&\displaystyle=-\frac{\pi}{A}\sum_{{\bm G}\neq 0}\mbox{K}(\alpha,G;h)\\
&\\
&\displaystyle \times \Im m \mbox{\Large{[}}\mbox{\large{(}}\sum_{i\in L_1}({\bm\mu}_i\cdot {\bm G})({\bm G}\cdot{\bm s}_i)\mbox{exp}(i{\bm G}\cdot{\bm s}_i)\mbox{\large{)}}\mbox{\large{(}}\sum_{j\in L_2}({\bm\mu}_j\cdot {\bm G})\mbox{exp}(-i{\bm G}\cdot{\bm s}_j)\mbox{\large{)}}\\
&\\
&\displaystyle -\mbox{\large{(}}\sum_{i\in L_1}({\bm\mu}_i\cdot {\bm G})\mbox{exp}(i{\bm G}\cdot{\bm s}_i)\mbox{\large{)}}\mbox{\large{(}} \sum_{j\in L_2}({\bm\mu}_j\cdot {\bm G})({\bm G}\cdot{\bm s}_j)\mbox{exp}(-i{\bm G}\cdot{\bm s}_j)\mbox{\large{)}}\mbox{\Large{]}}
\end{array}
\end{equation}
with functions $I$, $J$ and $K$ defined in Eq.(\ref{B5}).\\
The contributions to the normal component of the stress tensor are given by
\begin{equation}
\label{C9}
\begin{array}{ll}
\displaystyle  \left.\frac{\partial}{\partial z}E_{{\bm G}\neq 0}^{(1,inter)}\right|_{z=h}&\displaystyle =-\frac{\pi }{A}\sum_{{\bm G}\neq 0}\mbox{\Large{[}}  G^2\mbox{J}(\alpha,G;h)+\frac{4\alpha^3 h}{\sqrt{\pi}} \mbox{Q}(\alpha,G;h)\mbox{\Large{]}}\\
&\\
&\displaystyle\times\Re e\mbox{\Large{[}} \mbox{\large{(}}\sum_{i\in L_1} \mu_i^{z}\exp(i\bm{G}\cdot\bm{s}_i)\mbox{\large{)}} \mbox{\large{(}}\sum_{j\in L_2} \mu_j^{z}\exp(-i\bm{G}\cdot\bm{s}_j)\mbox{\large{)}}\mbox{\Large{]}}
\end{array}
\end{equation}\\
\begin{equation}
\label{C10}
\begin{array}{ll}
\displaystyle  \left.\frac{\partial}{\partial z}E_{{\bm G}\neq 0}^{(2,inter)}\right|_{z=h}&\displaystyle =\frac{\pi }{A}\sum_{{\bm G}\neq 0}\mbox{\Large{[}}G^2 \mbox{K}(\alpha,G;h)  -\frac{2\alpha}{\sqrt{\pi}} \mbox{P}(\alpha,G;h)\mbox{\Large{]}}\\
&\\
&\displaystyle\times\Im m\mbox{\Large{[}}\mbox{\large{(}}\sum_{i\in L_1} (\bm{\mu}_i\cdot\bm{G})\exp(i\bm{G}\cdot\bm{s}_i)\mbox{\large{)}}\mbox{\large{(}}\sum_{j\in L_2} \mu_j^{z}\exp(-i\bm{G}\cdot\bm{s}_j)\mbox{\large{)}} \\
&\\
&\displaystyle +\mbox{\large{(}}\sum_{i\in L_1} \mu_i^{z}\exp(i\bm{G}\cdot\bm{s}_i)\mbox{\large{)}} \mbox{\large{(}}\sum_{j\in L_2} (\bm{\mu}_j\cdot\bm{G})\exp(-i\bm{G}\cdot\bm{s}_j)\mbox{\large{)}}\mbox{\Large{]}}
\end{array}
\end{equation}\\
\begin{equation}
\label{C11}
\begin{array}{ll}
\displaystyle  \left.\frac{\partial}{\partial z}E_{{\bm G}\neq 0}^{(3,inter)}\right|_{z=h}&\displaystyle =\frac{\pi }{A}\sum_{{\bm G}\neq 0} \mbox{J}(\alpha,G;h)\\
&\\
&\displaystyle\times  \Re e\mbox{\Large{[}} \mbox{\large{(}}\sum_{i\in L_1} (\bm{\mu}_i\cdot\bm{G})\exp(i\bm{G}\cdot\bm{s}_i)\mbox{\large{)}} \mbox{\large{(}}\sum_{j\in L_2} (\bm{\mu}_j\cdot\bm{G})\exp(-i\bm{G}\cdot\bm{s}_j)\mbox{\large{)}}\mbox{\Large{]}}.
\end{array}
\end{equation}
The function $Q(\alpha,G;h)$ is obtained from the derivative of $J$, i.\ e.\,
\begin{equation}
\label{C12}
\displaystyle \mbox{Q}(\alpha,G;h) =2\exp\mbox{\large{(}}-\frac{G^2}{4\alpha^2}\mbox{\large{)}}\exp(-\alpha^2 h^2).
\end{equation}
Finally,
\begin{equation}
\label{C13}
\displaystyle  \left.\frac{\partial}{\partial z}E_{{\bm G}= 0}^{(inter)}\right|_{z=h}=-2\alpha^2 h E_{{\bm G}= 0}^{(inter)}
\end{equation}
with $E_{{\bm G}= 0}^{(inter)}$ given by Eq.(\ref{B6}).

\newpage
{\Large\bf List of Tables}\\[0.2in]

\normalsize{\bf Table \ref{tb:coef} :} Definitions of the projections of intralayer and interlayer correlation functions computed in the present work.\\[0.1in]
 
\normalsize{\bf Table \ref{tb:varh} :} Average energies and pressures for the bilayer system for $\rho=0.7$ and several values of $h$. The numbers in brackets give the accuracy on the last digit of the averages. $a$ is the width of the one-body orientational distribution obtained by fitting the MC histograms.  $\beta U_{dd}/N$, $\beta U^{intra}/N$  and $\beta U^{inter}/N$ denote, respectively, the averages of total, intralayer  and interlayer dipolar energies. $P_{zz}^{(dd)}$ is the average normal force by unit area as defined by Eq.(\ref{press4}). $\Pi_{T}^{(dd)}$ is the average of the dipolar contribution to the lateral pressure computed with Eq.(\ref{press1}) and $\Pi_{T}^{(HS)}$ is the hard sphere contribution computed from the contact value of the pair distribution function Eq.(\ref{press3}). $\tilde{\eta}= -2\rho kT - 2\Pi_{T}^{(HS)} - \Pi_{T}^{(dd)}$ is the surface stress as defined in Eq.(\ref{BB15}).\\[0.1in]

\normalsize{\bf Table \ref{tb:Ufixh} :} Average energies for the bilayer system for several values of $\rho$ and $\mu$ for $h=1.05$. Notations are the same as in Table \ref{tb:varh}.\\[0.1in]

\normalsize{\bf Table \ref{tb:Pfixh} :} Average pressures for the bilayer system for several values of $\rho$ and $\mu$ for $h=1.05$. Notations are the same as in Table \ref{tb:varh}.\\[0.1in]

\newpage
{\Large\bf List of Figures}\\[0.2in]

\normalsize{\bf Figure \ref{fig.1}:} Orientational distribution functions of dipolar moments in a bilayer of dipolar hard spheres. Symbols denote MC data and solid lines are fits using Eq.(\ref{fit1}). (a) Results at selected values of $\rho$ and $\mu$ at $h=1.05$. (b) Variation with layer separation $h$ for $\rho = 0.7$ and $\mu=2.00$. \\[0.1in]

\normalsize{\bf Figure \ref{fig.2}:} Average energies (a) and normal pressures (b) as a function of $h$ for $\rho=0.7$ and $\mu=1$ and 2. The symbols denote MC data and the lines are fits to the data using Eqs.(\ref{fit2}) and (\ref{fit3}), respectively. The fitting parameters for $\mu$=1 are $e_0= 0.16\pm 0.01$, $e_1= 0.045\pm 0.002$ and $f_0= 0.46\pm 0.01$, $f_1= 0.13 \pm 0.01 $. For $\mu$=2 they are  $e_0=0.31\pm 0.01$, $e_1= 0,28\pm 0.01$ and $f_0= 0.68\pm 0.03$, $f_1= 1.3\pm 0.1$. \\[0.1in]

\normalsize{\bf Figure \ref{fig.3}:} (a) Lateral pressure as a function of the intralayer energy
    per unit area. Symbols are data from Tables III and IV for densities
    $\rho=0.3-0.7$, dipole strengths $\mu=1.0-2.5$ and $h=1.05$ ; the
    straight line is given by Eq.(\ref{eos1}). (b) Surface stress as
    function of dipole strength for $\rho=0.3-0.7$ and
    $h=1.05$. Symbols are data from Table IV and lines are given by
    Eq.(\ref{fit4}) with $g_1(a_1 ;\rho,\mu)=a_1\rho^2\mu^4/(1+\mu^2)$
    ($a_1\sim 2.7$) and $\Pi_T^{(HS)}(\rho,0)$ given by the equation of
    state of hard disks ref.\ \cite{Santos:95}.\\[0.1in]

\normalsize{\bf Figure \ref{fig.4}:} Intralayer angle averaged pair distribution function (a) $g_{intra}^{000}(s)$ and angular projections (b) $h_{intra}^{klm}(s)$ for a bilayer of dipolar hard spheres at $\rho=0.7$, $h=1.05$ for $\mu=1.0$ (black) and $\mu=2.0$ (red -grey hue).\\[0.1in]

\normalsize{\bf Figure \ref{fig.5}:} Interlayer angle averaged pair distribution function $g_{inter}^{000}(s)$ and angular projections $h_{inter}^{klm}(s)$ of the pair distribution functions $g_{inter}(12)$ for the DHS bilayer at $\rho=0.3$ and $h=1.05$ for several values of $\mu$. (a) $g_{inter}^{000}(s)$ ; (b) $h_{inter}^{110}(s)$ ; (c)  $h_{inter}^{112}(s)$, (d)  $h_{inter}^{220}(s)$.\\[0.1in]

\normalsize{\bf Figure \ref{fig.6}:} Same as Fig. \ref{fig.5} but for $\rho=0.7$.\\[0.1in]

\normalsize{\bf Figure \ref{fig.7}:} Snapshots of  bilayer configurations of particles at
    $\mu=2.0$ ((a),(b)) and $\mu=2.50$ ((c),(d)) for  $h=1.05$ ;
    snapshots (a) and (c) are for
    $\rho=0.3$ ($N=1058$); snapshots (b) and (d) for $\rho=0.7$
    ($N=1024$). Particles in 
    different layers are represented by  different colours. The HS
    cores are represented by circles of diameter $\sigma=1$ and the
    directions of dipole moments by arrows.\\[0.1in]

\normalsize{\bf Figure \ref{fig.8}:} Bilayer configurations of the $2 \times 1600$ particle system  at
    $\rho=0.9$,  $\mu=2$ and  $h=1.05$  at different intervals of the MC
    simulation; (a) snapshot after 500 cycles, (b) $0.26 \times 10^6$
    cycles, (c) $1.75 \times 10^6$ cycles, (d) result for $2 \times 576$
    particles 
    after  $2.6 \times 10^6$ cycles. For clarity only the particle
    arrangements in  one layer are shown in  (a)-(c). The
    arrows denote the projections of the dipole moments on the layer
    plane. Thus dipoles perpendicular to the layer appear as dots.\\[0.1in] 

\normalsize{\bf Figure \ref{fig.9}:} Snapshots of  bilayer configurations of particles at close
    packing.  (a) square lattice  ($\rho=1$,  $\mu=2$, $h=1.05$, N=3200);
    (b) hexagonal lattice  ($\rho=1.15$,  $\mu=2$, $h=1.05$, N=2400). 
     The particles in the two layers are on top of each other. The
    arrows denote the projections of the dipole moments on the layer
    plane. The two layers are shown separately.\\[0.1in] 
    
\newpage
\begin{table}
\begin{center}
\footnotesize
\begin{tabular}{cc|c}
\hline
\hline
& & \\
 $(l_1,l_2,l)$ & $\tilde{\Phi}^{l_1 l_2 l}$ &  Intralayer-Interlayer \\
    &  & functions  \\
& &\\
\hline
&  &\\
(0,0,0) & 1 &  $g_{intra,inter}^{000}(s)=<g_{intra,inter}(12)>_{\hat{\bm{\mu}}_1\hat{\bm{\mu}}_2}$  \\ 
&  &\\
(1,1,0) & $\hat{\bm{\mu}}_1\cdot\hat{\bm{\mu}}_2$ &  $h_{intra,inter}^{110}(s)=3<g_{intra,inter}(12)\tilde{\Phi}^{110}(12)>_{\hat{\bm{\mu}}_1\hat{\bm{\mu}}_2}$ \\
&  &\\
(1,1,2) & $3(\hat{\bm{\mu}}_1\cdot\hat{\bm{r}})(\hat{\bm{\mu}}_2\cdot\hat{\bm{r}})-\hat{\bm{\mu}}_1\cdot\hat{\bm{\mu}}_2$ & $h_{intra,inter}^{112}(s)=\frac{3}{2}<g_{intra,inter}(12)\tilde{\Phi}^{112}(12)>_{\hat{\bm{\mu}}_1\hat{\bm{\mu}}_2}$ \\
&  &\\
(2,2,0) & $\frac{1}{2}(3(\hat{\bm{\mu}}_1\cdot\hat{\bm{\mu}}_2)^2-1)$ & $h_{intra,inter}^{220}(s)= \frac{5}{2} <g_{intra,inter}(12)\tilde{\Phi}^{220}(12)>_{\hat{\bm{\mu}}_1\hat{\bm{\mu}}_2}$\\
&  &\\
\hline
\hline
\end{tabular}\\[2.5in]
\caption{Definitions of the projections of intralayer and interlayer correlation functions computed in the present work. }
\label{tb:coef}
\end{center}
\end{table}

\newpage
\begin{table}
\scriptsize
\begin{tabular}{c|ccccccccc}
\hline
\hline
&&&&&&&&&\\
$\mu$ & $h$ & $a$ & $\beta U_{dd}/N$ & $\beta U^{intra}/N$ & $\beta U^{inter}/N$ & $P_{zz}^{(dd)}$ & $\Pi_{T}^{(dd)}$ & $\Pi_{T}^{(HS)}$ &  $\tilde{\eta}$ \\
&&&&&&&&&\\
\hline
&&&&&&&&&\\
1.00 & 1.05 & 0.36 &-0.70(2) & -0.55(2) & -0.16(1)  & -0.44(4)  & -1.24(5) &  3.2(2) & -6.6(2)\\
     & 1.15 & 0.43 &-0.67(2) & -0.57(2) & -0.10(1)  & -0.26(3)  & -1.25(4) &  3.1(1) & -6.4(1)\\
     & 1.25 & 0.48 &-0.65(2) & -0.58(2) & -0.07(1)  & -0.17(2)  & -1.26(4) &  3.1(1) & -6.3(1)\\
     & 1.35 & 0.52 &-0.64(2) & -0.59(2) & -0.05(1)  & -0.11(1)  & -1.27(4) &  3.1(1) & -6.3(1)\\
     & 1.45 & 0.52 &-0.63(2) & -0.59(2) & -0.04(1)  & -0.07(1)  & -1.28(4) &  3.2(1) & -6.5(1)\\
     & 1.55 & 0.53 &-0.62(2) & -0.60(2) & -0.03(1)  & -0.05(1)  & -1.27(3) &  3.1(1) & -6.3(1)\\
     & 1.65 & 0.57 &-0.62(2) & -0.60(2) & -0.021(5) & -0.03(1)  & -1.27(4) &  3.1(1) & -6.3(1)\\
     & 1.80 & 0.55 &-0.62(2) & -0.60(2) & -0.015(4) & -0.02(1)  & -1.29(4) &  3.1(1) & -6.3(1)\\
     & 1.95 & 0.57 &-0.62(2) & -0.61(2) & -0.011(3) & -0.015(5) & -1.29(4) &  3.1(1) & -6.3(1)\\
     & 2.10 & 0.57 &-0.62(2) & -0.61(2) & -0.008(3) & -0.010(4) & -1.28(4) &  3.2(1) & -6.5(1)\\  
     & 2.40 & 0.57 &-0.61(2) & -0.61(2) & -0.005(2) & -0.005(3) & -1.27(4) &  3.1(1) & -6.3(1)\\                        
     & 3.00 & 0.58 &-0.61(2) & -0.61(2) & -0.003(2) & -0.002(1) & -1.26(4) &  3.2(1) & -6.5(1)\\
&&&&&&&&&\\
\hline
&&&&&&&&&\\
2.00 & 1.01 & 3.8 & -6.3(1) & -5.7(1) & -0.56(4)  & -1.8(1)  & -12.2(1)  & 6.5(3) & -2.2(4)\\
     & 1.05 & 4.2 & -6.3(1) & -5.8(1) & -0.42(3)  & -1.3(1)  & -12.5(1)  & 6.8(3) & -2.5(4)\\
     & 1.10 & 4.6 & -6.3(1) & -5.9(1) & -0.32(3)  & -0.9(1)  & -12.7(1)  & 6.8(3) & -2.3(4)\\
     & 1.15 & 4.8 & -6.3(1) & -6.1(1) & -0.24(2)  & -0.6(1)  & -12.8(1)  & 7.0(3) & -2.6(4)\\
     & 1.25 & 5.2 & -6.3(1) & -6.1(1) & -0.16(2)  & -0.32(3) & -13.0(1)  & 7.0(3) & -2.4(4)\\
     & 1.35 & 5.4 & -6.3(1) & -6.2(1) & -0.11(1)  & -0.20(2) & -13.1(1)   & 7.1(3) & -2.5(4)\\
     & 1.45 & 5.5 & -6.3(1) & -6.2(1) & -0.08(1)  & -0.13(2) & -13.1(1)   & 7.1(3) & -2.5(4)\\
     & 1.55 & 5.6 & -6.3(1) & -6.21(5) & -0.06(1) & -0.09(2) & -13.2(1)   & 7.1(3) &  -2.4(4)\\
     & 1.65 & 5.6 & -6.3(1) & -6.23(5) & -0.05(1) & -0.06(1) & -13.2(1)   & 7.1(3) &  -2.4(4)\\
     & 1.80 & 5.7 & -6.3(1) & -6.25(5) & -0.03(1) & -0.03(1) & -13.1(1)   & 7.1(3) &  -2.5(4)\\
     & 1.95 & 5.7 & -6.3(1) & -6.28(5) & -0.02(1) & -0.03(1) & -13.3(1)   & 7.1(3) &  -2.3(4)\\
     & 2.10 & 5.7 & -6.3(1) & -6.28(5) & -0.019(5)& -0.019(5)& -13.2(1)  & 7.1(3) &  -2.4(4)\\ 
&&&&&&&&&\\
\hline
\hline
\end{tabular}\\[0.5in]
\caption{Average energies and pressures for the bilayer system for $\rho=0.7$ and several values of $h$. The numbers in brackets give the accuracy on the last digit of the averages. $a$ is the width of the one-body orientational distribution obtained by fitting the MC histograms.  $\beta U_{dd}/N$, $\beta U^{intra}/N$  and $\beta U^{inter}/N$ denote, respectively,  the averages of total, intralayer  and interlayer dipolar energies. $P_{zz}^{(dd)}$ is the average normal force by unit area as defined by Eq.(\ref{press4}). $\Pi_{T}^{(dd)}$ is the average of the dipolar contribution to the lateral pressure computed with Eq.(\ref{press1}) and $\Pi_{T}^{(HS)}$ is the hard sphere contribution computed from the contact value of the pair distribution function Eq.(\ref{press3}). $\tilde{\eta}= -2\rho kT - 2\Pi_{T}^{(HS)} - \Pi_{T}^{(dd)}$ is the surface stress as defined in Eq.(\ref{BB15}).}
\label{tb:varh}
\end{table}

\newpage
\begin{table}
\scriptsize
\begin{tabular}{cccccc|cccccc}
\hline
\hline
&&&&&&&&&&&\\
$\mu$ & $\rho$ & $\beta U_{dd}/N$ & $\beta U^{intra}/N$ & $\beta U^{inter}/N$ & $a$ &  $\mu$ & $\rho$ & $\beta U_{dd}/N$ &  $\beta U^{intra}/N$ & $\beta U^{inter}/N$ & $a$  \\
&&&&&&&&&&&\\
\hline
&&&&&&&&&&&\\
1.00 & 0.3 & -0.29(1) & -0.19(1) & -0.10(1) & 0.07 & 2.00 & 0.3 & -4.9(1)   & -4.3(1)   & -0.52(3) & 3.1  \\ 
        & 0.4 & -0.39(2) & -0.26(2) & -0.12(1) & 0.12 &         & 0.4 & -5.2(1)   & -4.7(1)   & -0.50(3) & 3.4  \\ 
        & 0.5 & -0.49(2) & -0.35(2) & -0.14(1) & 0.18 &         & 0.5 & -5.6(1)   & -5.1(1)   & -0.48(4) & 3.6   \\
        & 0.6 & -0.59(2) & -0.44(2) & -0.15(1) & 0.25 &         & 0.6 & -5.8(1)   & -5.4(1)   & -0.46(3) & 3.8  \\ 
        & 0.7 & -0.70(2) & -0.54(2) & -0.16(1) & 0.35 &         & 0.7 & -6.3(1)   & -5.8(1)   & -0.42(4) & 4.2   \\ 
1.25 & 0.3 & -0.68(2) & -0.46(2) & -0.22(2) & 0.19 & 2.25 & 0.3 & -8.1(1)   & -7.8(1)   & -0.34(3) & 6.3  \\ 
        & 0.4 & -0.87(3) & -0.62(3) & -0.25(2) & 0.31 &         & 0.4 & -8.3(1)   & -7.9(1)   & -0.35(3) & 6.3  \\ 
        & 0.5 & -1.06(3) & -0.80(3) & -0.27(2) & 0.44 &         & 0.5 & -8.4(1)   & -8.0(1)   & -0.38(3) & 6.2  \\ 
        & 0.6 & -1.26(3) & -0.99(3) & -0.27(2) & 0.60 &         & 0.6 & -8.6(1)   & -8.2(1)   & -0.39(4) & 6.2  \\ 
        & 0.7 & -1.46(3) & -1.19(3) & -0.27(2) & 0.78 &         & 0.7 & -8.9(1)   & -8.6(1)   & -0.39(3) & 6.4   \\ 
1.50 & 0.3 & -1.38(4) & -1.01(4) & -0.37(2) & 0.51 & 2.50 & 0.3 & -11.7(1) & -11.5(1) & -0.20(2) & 9.9  \\
        & 0.4 & -1.70(4) & -1.29(4) & -0.41(2) & 0.71 &         & 0.4 & -11.7(1) & -11.5(1) & -0.25(2) & 9.6  \\ 
        & 0.5 & -2.00(4) & -1.59(4) & -0.41(2) & 0.95 &         & 0.5 & -11.8(1) & -11.5(1) & -0.28(3) & 9.4  \\ 
        & 0.6 & -2.29(5) & -1.89(5) & -0.39(3) & 1.2   &         & 0.6 & -11.9(1) & -11.6(1) & -0.32(1) & 9.2  \\
        & 0.7 & -2.60(5) & -2.22(5) & -0.37(3) & 1.5   & 	       & 0.7 & -12.2(1) & -11.9(1) & -0.34(2) & 9.2  \\
1.75 & 0.3 & -2.65(5) & -2.13(5) & -0.52(3) & 1.3   & & & & & &   \\ 
        & 0.4 & -3.1(1)   & -2.5(1)   & -0.52(3) & 1.6   & & & & & &   \\ 
        & 0.5 & -3.4(1)   & -2.9(1)   & -0.50(4) & 1.9   & & & & & &   \\ 	
        & 0.6 & -3.8(1)   & -3.3(1)   & -0.46(3) & 2.3   & & & & & &   \\ 
        & 0.7 & -4.2(1)   & -3.8(1)   & -0.42(3) & 2.6   & & & & & &   \\ 

&&&&&&&&&&&\\
\hline
\hline
\end{tabular}\\[0.5in]
\caption{Average energies for the bilayer system for several values of $\rho$ and $\mu$ for $h=1.05$. Notations are the same as in Table \ref{tb:varh}.}
\label{tb:Ufixh}
\end{table}

\newpage
\begin{table}
\scriptsize
\begin{center}
\begin{tabular}{cccccc|cccccc}
\hline
\hline
&&&&&&&&&&&\\
$\mu$ & $\rho$ &  $P_{zz}^{(dd)}$ & $\Pi_{T}^{(dd)}$ & $\Pi_{T}^{(HS)}$ & $\tilde{\eta}$ & $\mu$ & $\rho$ &  $P_{zz}^{(dd)}$ & $\Pi_{T}^{(dd)}$ & $\Pi_{T}^{(HS)}$ & $\tilde{\eta}$\\
&&&&&&&&&&&\\
\hline
&&&&&&&&&&&\\
1.00 & 0.3 & -0.12(1) & -0.20(1) & 0.26(1) & -0.92(2) & 2.00  & 0.3 & -0.62(4) & -4.0(1)   & 1.8(1)   & -0.2(2)  \\
	 & 0.4 & -0.20(2) & -0.36(2) & 0.55(3) & -1.54(5) &          & 0.4 & -0.80(5) & -5.9(1)   & 2.7(1)   & -0.3(2)  \\
	 & 0.5 & -0.28(3) & -0.59(3) & 1.02(5) & -2.4(1)   &          & 0.5 & -1.0(1)   & -7.9(1)   & 3.7(2)   & -0.5(2)  \\
	 & 0.6 & -0.36(3) & -0.87(3) & 1.8(1)   & -3.9(1)   &          & 0.6 & -1.2(1)   & -10.0(1) & 5.0(2)   & -1.2(2)  \\
	 & 0.7 & -0.44(4) & -1.24(4) & 3.1(2)   & -6.4(1)   &          & 0.7 & -1.28(1) & -12.5(1) & 6.8(3)   & -2.5(3)  \\
1.25 & 0.3 & -0.26(2) & -0.47(2) & 0.34(2) & -0.81(4) & 2.25  & 0.3 & -0.43(4) & -7.0(1)   & 3.1(2)   & 0.2(2)  \\
	 & 0.4 & -0.40(3) & -0.84(3) & 0.69(3) & -1.34(5) &          & 0.4 & -0.59(5) & -9.7(1)   & 4.4(2)   & -0.1(2)  \\
	 & 0.5 & -0.54(4) & -1.31(4) & 1.23(6) & -2.1(1)   &          & 0.5 & -0.8(1)   & -12.2(1) & 5.6(3)   & 0.0(3)  \\
	 & 0.6 & -0.67(5) & -1.91(5) & 2.1(1)   & -3.5(1)   &          & 0.6 & -1.1(1)   & -15.0(1) & 7.0(4)   & -0.2(4)  \\
	 & 0.7 & -0.78(5) & -2.6(1)   & 3.5(2)   & -5.8(1)   &          & 0.7 & -1.26(3) & -18.2(2) & 9.1(5)   & -1.4(5)  \\	 	 	 
1.50 & 0.3 & -0.45(3) & -1.01(3) & 0.50(3) & -0.59(5) & 2.50  & 0.3 & -0.32(3) & -10.3(1) & 4.3(2)   & 1.1(2)  \\
	 & 0.4 & -0.66(4) & -1.69(5) & 0.98(5) & -1.1(1)   &          & 0.4 & -0.54(4) & -13.8(1) & 6.3(3)   & 0.4(3)  \\
	 & 0.5 & -0.83(5) & -2.55(5) & 1.6(1)   & -1.7(1)   &          & 0.5 & -0.76(5) & -17.2(1) & 7.6(4)   & 1.0(4)  \\
	 & 0.6 & -1.0(1)   & -3.6(1)   & 2.6(1)   & -2.8(2)   &          & 0.6 & -1.0(1)   & -20.8(1) & 9.4(5) & 0.8(5)  \\
	 & 0.7 & -1.07(1) & -4.9(1)   & 4.1(2)   & -4.7(2)   &          & 0.7 & -1.5(1)   & -24.9(1) & 12.1(5) & -0.7(5) \\
1.75 & 0.3 & -0.62(4) & -2.05(5) & 0.85(4) &  -0.3(1)   &  &  &  &  &  &    \\
	 & 0.4 & -0.83(5) & -3.3(1)   & 1.6(1)   &  -0.7(2)   &  &  &  &  &  &    \\
	 & 0.5 & -1.0(1)   & -4.6(1)   & 2.4(1)   &  -1.2(2)   &  &  &  &  &  &    \\
	 & 0.6 & -1.1(1)   & -6.2(1)   & 3.5(2)   &  -2.0(3)   &  &  &  &  &  &    \\
	 & 0.7 & -1.23(5) & -8.1(1)   & 5.2(3)   &  -3.7(3)   &  &  &  &  &  &    \\
&&&&&&&&&&&\\
\hline
\hline
\end{tabular}\\[0.5in]
\end{center}
\caption{Average pressures for the bilayer system for several values of $\rho$ and $\mu$ for $h=1.05$. Notations are the same as in Table \ref{tb:varh}.}
\label{tb:Pfixh}
\end{table}

\newpage

\newpage

\newpage

\newpage

\newpage

\newpage

\newpage

\newpage

\newpage

\end{document}